\documentclass[]{aastex} 

\usepackage[fleqn]{amsmath}
\usepackage{amssymb}
\usepackage{graphicx}
\usepackage{txfonts}
\usepackage{natbib}
\usepackage{subfigure}
\usepackage{url}
\usepackage{multirow}
\usepackage{gensymb}
\usepackage[usenames,dvipsnames]{xcolor}
\usepackage[version=3]{mhchem}

\newcommand{\myemaila}{cwalsh@strw.leidenuniv.nl}
\newcommand{\myemailb}{c.walsh1@leeds.ac.uk}

\shorttitle{ALMA OBSERVATIONS OF THE HD~97048 DISK}
\shortauthors{Catherine Walsh et al.}

\begin{document}

\title{ALMA REVEALS THE ANATOMY OF THE mm-SIZED DUST AND MOLECULAR GAS IN THE HD~97048 DISK}

\author{Catherine Walsh\altaffilmark{1,2}, 
Attila Juh\'{a}sz\altaffilmark{3}, 
Gwendolyn Meeus\altaffilmark{4}, 
William~R.~F.~Dent\altaffilmark{5},
Luke Maud\altaffilmark{1},
Yuri Aikawa\altaffilmark{6}, 
Tom~J.~Millar\altaffilmark{7}, 
Hideko Nomura \altaffilmark{8}} 

\altaffiltext{1}{Leiden Observatory, Leiden University, P.O.~Box 9531, 2300~RA Leiden, The Netherlands}
\altaffiltext{2}{School of Physics and Astronomy, University of Leeds, Leeds LS2 9JT, UK}
\altaffiltext{3}{Institute of Astronomy, University of Cambridge, Madingley Road, Cambridge CB3 0HA, UK}
\altaffiltext{4}{Departamento de F\'{i}sica Te\'{o}rica, Universidad Autonoma de Madrid, Campus Cantoblanco, E-28049 Madrid, Spain}
\altaffiltext{5}{Joint ALMA Observatory (JAO), Alonso de C\'{o}rdova 3107, Vitacura, Santiago, Chile}
\altaffiltext{6}{Center for Computer Sciences, University of Tsukuba, 305-8577 Tsukuba, Japan}
\altaffiltext{7}{School of Mathematics and Physics, Queen's University Belfast, University Road, Belfast BT7 1NN, UK}
\altaffiltext{8}{Department of Earth and Planetary Science, Tokyo Institute of Technology, 2-12-1 Ookayama, Meguro-ku, 152-8551 Tokyo, Japan}

\email{\myemaila;\myemailb}
 
\begin{abstract}

Transitional disks show a lack of excess emission at 
infrared wavelengths due to a large dust cavity, 
that is often corroborated by spatially resolved 
observations at $\sim$mm wavelengths. 
We present the first spatially resolved $\sim$mm-wavelength images of the disk 
around the Herbig Ae/Be star, HD~97048.  
Scattered light images show that the disk extends to $\approx640$~au.  
The ALMA data reveal a circular-symmetric dusty disk 
extending to $\approx350$~au, and a molecular disk traced in 
CO $J=3$-2 emission, extending to $\approx750$~au.  
The CO emission arises from a flared layer with an 
opening angle $\approx30\degree-40\degree$.  
HD~97048 is another source for which the large ($\sim$~mm-sized) dust grains 
are more centrally concentrated than the small ($\sim\mu$m-sized) grains and 
molecular gas, likely due to radial drift.
The images and visibility data modeling suggest a decrement in continuum emission 
within $\approx50$~au, consistent with the cavity size determined from 
mid-infrared imaging ($34\pm4$~au).  
The extracted continuum intensity profiles show ring-like structures 
with peaks at $\approx 50$, 150, and 300~au, with associated gaps at 
$\approx100$ and 250~au. This structure should be confirmed   
in higher-resolution images (FWHM~$\approx 10-20$~au).  
These data confirm the classification of HD~97048 as a transitional disk 
that also possesses multiple ring-like structures in the dust continuum 
emission.  
Additional data are required at multiple and well-separated frequencies 
to fully characterize the disk structure, and thereby constrain 
the mechanism(s) responsible for sculpting the HD~97048 disk.
\end{abstract}


\section{INTRODUCTION}
\label{introduction}

Protoplanetary disks are the sites of planetary system formation.  
So-called {\em transitional disks} are considered a particular class of 
protoplanetary disk which have substantial dust cavities  
\cite[or gaps, see the recent review by][p.~497]{espaillat14}.  
The origin of dust cavities in transitional disks is 
much debated in the literature.  
Theories range from photoevaporation of dust and gas by the 
central star \cite[see, e.g.,][p.~475]{alexander14} to the development of 
{\em dead zones}, regions of low ionization which impede 
the angular momentum and mass transport leading to the build-up of 
material in a ring-like structure \citep[][]{regaly13,flock15}.  
A third theory is that unseen massive 
planets or companions in the disk create steep pressure gradients 
in the gas, which shepherd large dust grains into rings (so-called ``dust traps''), 
generating the appearance of a large cavity when imaged at (sub-)mm wavelengths 
\citep[][]{andrews11,pinilla12b}.  
Distinguishing between theories requires complementary 
observations of the molecular gas and imaging of the emission from both 
small ($\sim\mu$m) and large ($\sim$mm) dust grains 
\citep[see Table 2 in][p.~497]{espaillat14}.  
Cavities in both dust (small and large) and gas are likely created by 
photoevaporation whereas those created by pressure bumps (possibly 
triggered by planets) are devoid of large (mm) grains only, 
and residual gas and small dust grains may remain 
\citep[due to dust filtration, see, e.g.,][]{rice06}.  

Transitional disks were originally identified via the lack 
of near- to mid-infrared (IR) emission in the spectral energy distributions (SEDs) of 
disk-hosting stars, indicating a depletion in warm dust and 
hence, the presence of an inner dust cavity \citep[see, e.g.,][]{strom89}.  
Several transitional disks encompass Herbig Ae/Be (HAeBe) stars 
\citep[$T_\mathrm{eff}$, $\sim$10,000~K,][]{waters98}.  
HAeBe star-disk systems have been further classified by the shape of the SEDs 
in the mid-IR into either Group I or Group II disks; the former have a flared 
structure (driven by absorption of stellar UV photons) and the latter are 
considered to be ``flat'' disks \citep{meeus01}, within which the 
dust has grown to mm-sizes, has become decoupled from the gas, and 
has settled to the midplane \citep{dullemond04}. 
This scenario of the gradual depletion of small dust grains
which absorbs the UV radiation necessary to trigger the flared structure has led to 
speculation in the literature that Group II disks 
may be a later evolutionary state of Group I disks \citep{dullemond04}.  
The advent of interferometric imaging across the wavelength range of interest 
has thrown doubt on this rather intuitive evolutionary scenario.  
High spatial-resolution imaging at long ($\gtrsim$~mm) wavelengths with 
ALMA, JVLA, and ATCA have revealed 
that grain growth can already be advanced in Group I disks 
($\beta\lesssim 1$, where $F_\nu \propto \nu^{\beta + 2}$), e.g., 
HD~142527 \citep[][]{casassus13,casassus15}, 
IRS~48 \citep[][]{vandermarel13,vandermarel15}, 
and HD~100546 \citep[][]{pineda14,walsh14,wright15}. 
It should be noted that the value of $\beta$ is also sensitive 
to grain composition \citep{draine06}. 
Recent mid-IR imaging also suggests that all Group I 
disks are transitional in nature, and the resulting flared structure 
is therefore linked to the presence of the inner cavity 
\citep[][]{honda12,maaskant13}.    
Assuming the theory that massive planets are responsible for 
generating such cavities, perhaps primordial protoplanetary disks 
follow one of two evolutionary paths with the division into Group I 
or Group II, dependent upon the formation of a massive 
companion (and associated cavity) early in the disk lifetime 
\citep[see, e.g.,][]{currie09}.
The picture is further complicated by recent 
mid-IR interferometric observations which suggest that several 
Group II HAeBe disks also exhibit evidence of 
inner cavities, albeit smaller than those typically seen in 
Group I disks \citep{menu15}. 
These authors hypothesize that the Group II objects are  
younger, and the evolution into a Group I disk follows the 
formation of an inner cavity. 

It is clear that complementary observations across the wavelength 
range of interest, from near-IR to (sub-)mm, are necessary to help further 
elucidate the evolutionary paths and states of HAeBe disks and the physical 
origin of the cavities observed therein.  
We present spatially resolved ALMA Cycle 0 observations 
of the Group I HAeBe star-disk system, HD~97048.  
The observations reveal for the first time the spatial distribution of the large 
($\sim$~mm-sized) dust grains and molecular gas in this otherwise 
well-studied source.    
In Sections~\ref{hd97048}-\ref{discussion} 
we describe the source, the observations, 
present our results, and discuss the implications, respectively. 
    
\section{THE SOURCE: HD~97048}
\label{hd97048}

HD~97048 is a Herbig~Ae/Be star with spectral type B9--A0 ($T_\star\approx$~10,000~K) 
and a mass of 2.5~$M_\odot$, located in the Chameleon~I star-forming region
at a distance of $\approx$160~pc \citep{vandenancker98,vanleeuwen07}.     
To date, the HD~97048 star-disk system has been well studied at short wavelengths only.  
Early observations from 2.0 to 4.0~$\mu$m with the Anglo-Australian 
Telescope revealed strong emission features between 3.3 and 3.6~$\mu$m \citep{blades80}.    
Later observations with the Infrared Space Observatory (ISO) 
revealed additional emission at 6.2, 7.7, 8.6, and 11.3~$\mu$m, 
attributed to polyaromatic hydrocarbons \citep[PAHs,][]{gurtler99,vankerckhoven02}.  
Mid-IR imaging showed extended PAH emission (out to several hundred au) 
in line with that predicted by flared protoplanetary disk models 
\citep{vanboekel04,lagage06,doucet07,marinas11}.  
The lack of any silicate features in the IR spectra suggested that the 
opacity is dominated by small carbonaceous dust grains, a property  
supported by more recent data from {\em Spitzer} \citep{acke10,juhasz10}.  
Scattered light images at optical wavelengths using the 
{\em Hubble Space Telescope}/Advanced Camera for Surveys revealed the 
radial extent of the small dust grains, 
{$\approx640$~au (assuming a source distance of 160~pc), and evidence of 
filamentary structures postulated to be spiral arms \citep{doering07}.  
PDI (polarimetric differential imaging) images 
with VLT/NACO in the $H$ and $K_s$ bands (1.66 and 2.18 $\mu$m, respectively) 
revealed the innermost dust \citep[$16-160$~au,][]{quanz12}. 
However, recent fits to the SED 
and azimuthally averaged mid-IR brightness profile ($Q$-band, 24.5~$\mu$m) 
unearthed the potential transitional nature of HD~97048 for the first time  
\citep{maaskant13}.  
The authors find that the SED and brightness profile are best fit with a 
dust cavity $34\pm4$~au in radius and also infer the presence of an 
inner, optically thick, dusty disk between 0.3 and 2.5~au.  

Regarding the gas, detection of the [OI] line at 6300~$\AA$ uncovered 
an inner rotating gas disk \citep[][]{acke06}. 
High-spectral resolution observations of CO 
rovibrational emission revealed a significant reservoir of 
warm molecular gas in Keplerian motion within $\approx100$~au 
\citep{vanderplas09,vanderplas15}. 
\citet{vanderplas09} also infer a cavity in the inner CO reservoir of 11~au,  
suggesting efficient photodissociation of CO by the harsh stellar 
radiation field.    
There is also evidence that HD~97048 hosts a gas-rich outer disk:   
\ce{H2} emission has been detected by numerous groups 
\citep{martin-zaidi07,martin-zaidi09,bary08,carmona11} 
allowing estimates on the spatial extent of the
warm gas (out to a few 10\,s of au). 
HD~97048 was also a target in two {\em Herschel} Open Time Key Programs 
dedicated to characterizing the gas in protoplanetary disks, 
GASPS (P.I., W. R. F. Dent) and DIGIT (P.I., N. J. Evans).   
\citet{meeus12,meeus13} and \citet{fedele13} reported PACS detections of 
line emission from [OI] at 63.18 and 145.53~$\mu$m and from 
[CII] at 157.75~$\mu$m.
Also detected were several high-$J$ rotational lines of 
CO (ranging from $J=15$-14 to $J=30$-29), multiple lines of OH, and 
two rotational transitions of \ce{CH+} ($J=6$-5 and 5-4).  
\citet{vanderwiel14} report additional detections ($>3\sigma$) of 
CO rotational line emission (from $J=9$-8 to $J=13$-12) with SPIRE.   

Despite being well-studied at shorter wavelengths, there 
is a lack of data at (sub-)mm wavelengths.  
\citet{henning94} detected continuum emission at 1.3~mm 
using the SEST~15m telescope measuring a flux of $451.5\pm34$~mJy.  
Follow-up mapping showed that the emission was compact, 
although at that time it had not yet been inferred that this source 
hosted a protoplanetary disk \citep{henning98}.  
More recent LABOCA (The Large Apex BOlometer CAmera) observations 
determined a flux of $2610\pm131$~mJy at 870~$\mu$m \citep{phillips10}.  
\citet{hales14} report emission from the
CO $J=3$-2 line (346~GHz) towards HD~97048 in an APEX survey of 
nearby T~Tauri and Herbig~Ae stars; however, they conclude that the line 
emission is heavily contaminated by emission from the background 
Chameleon I molecular cloud, based upon an assumption for the source velocity 
obtained from \citet{carmona11}.   
Very recently, \citet{kama16} conducted an APEX survey of 
[CI] $^{3}\mathrm{P}_{1}-^{3}\mathrm{P}_{0}$ (492.161~GHz), 
[CI] $^{3}\mathrm{P}_{2}-^{3}\mathrm{P}_{1}$ (809.342~GHz) and 
CO $J=6$-5 (691.473~GHz) line emission from a sample of nearby 
protoplanetary disks (33 sources in total).  
All three lines were detected toward HD~97048, with this source having 
the brightest intrinsic line emission compared with all other sources in the sample.
HD~97048 was also one of the brightest emitters in all far-IR emission lines 
in the {\em Herschel} PACS data \citep{meeus12,meeus13,fedele13}, making this one of the 
brightest known disks to date in sub-mm line emission.    
All data discussed thus far have been conducted with single-dish telescopes and thus 
are spatially unresolved.  
The ALMA data presented here are the first to spatially resolve both 
the dust emission and the molecular disk at (sub-)mm wavelengths.  

\section{OBSERVATIONS}
\label{observations}

HD~97048 was observed with 25 antennas in a compact configuration 
during ALMA Cycle~0 operations on 2012 December 14 
(program 2011.0.00863.S, P.I.~C.~Walsh).   
Baselines ranged from 15.1 to 402~m and 
the target was observed in seven spectral windows centered 
at 300.506, 301.286, 303.927, 344.311, 345.798, 346.998, 
and 347.331~GHz. 
The channel width was 122.070~kHz 
corresponding to 0.12 km~s$^{-1}$ at 302~GHz and 0.11 km~s$^{-1}$ at 346~GHz;  
hence, the spectral resolution, applying Hanning smoothing, 
was 0.24 and 0.22 km~s$^{-1}$, respectively,     
The total on-source time was 24~minutes 17~s for the first execution 
(at the lower frequency) 
and 23~minutes 59~s for the second execution (at the higher frequency).  
The quasars J0522-364 and J1147-6753 were used as bandpass and phase calibrators, 
respectively, and Callisto was used as amplitude calibrator 
for the data at $\approx302$~GHz.  
For the data at the higher frequency ($\approx346$~GHz), the quasar 3c279  
was used as bandpass and amplitude calibrator, and J1147-6753 was used as phase calibrator.  

Upon inspection of the delivered data at $\approx346$~GHz, it
was discovered that, due to the lack of a specified flux calibrator observation, the flux scaling had not 
been set in the initial calibration; hence, we recalibrated the raw data using 3c279 as the flux calibrator. 
3c279 had not been monitored regularly in Band 7 during the time of our observations with 
the nearest data taken more than two months either side of the date at which our data were taken. 
As a solution, we used the two Band 3 monitoring measurements at $\approx98$ and $\approx110$~GHz 
taken just three days before our observations to calculate the spectral index of 3c279 and thus bootstrap the 
Band 3 flux to our Band 7 data (using the ALMA Analysis Utilities\footnotemark~task, \texttt{GetALMAFlux}\footnotemark).  
This reported the spectral index for 3c279 as $0.73\pm0.01$ and the flux at 
Band 7 (346~GHz) as $8.86\pm0.51$~Jy. 
Although this corresponds to a flux uncertainty of $\approx5$\%, 
considering the variable spectral index of 3c279 (for instance, the next monitoring 
observations taken two months later indicate that the spectral index becomes shallower), 
we estimate a total error on the flux of the source at 346~GHz as closer to $\approx$15\%, but this 
could be underestimated.  

The data were self-calibrated using CASA version 4.3, during which 
several antennas were flagged reducing the total number of antennas to 22. 
We used the continuum data at each respective wavelength to self-calibrate 
all data (continuum and lines) before final imaging.  
We used a timescale of 20~s (corresponding to $\approx3$ times the integration step),  
and increased the dynamic range of the continuum data by a factor of $\approx20$.  
After removal of line-containing channels and 
edge channels, the total available continuum bandwidth amounted to  
1.39~GHz at 302~GHz ($993~\mu$m) and 1.86~GHz at 346~GHz ($866~\mu$m).  
Having two essentially independent observations at different 
frequencies allowed us to assess the reality of low-level features in the 
data.  

\footnotetext{\url{https://casaguides.nrao.edu/index.php?title=Analysis_Utilities}}
\footnotetext{\url{https://safe.nrao.edu/wiki/bin/view/ALMA/GetALMAFlux}}

\section{RESULTS}
\label{results}

\subsection{Continuum emission}
\label{continuum}

\subsubsection{Continuum images}
\label{images}

In Figure~\ref{figure1}, we display the continuum emission 
at 302 and 346~GHz imaged 
using the CLEAN\footnotemark~algorithm with Briggs weighting (robust~=~0.5, top panels).  
This resulted in synthesized beams of 0\farcs88$\times$0\farcs52 (-43\fdg9) 
at 302~GHz and 0\farcs69$\times$0\farcs43 (-23\fdg2) at 346~GHz.  
The elliptical beams are due to the relatively low decl.~of the 
source ($-77\degree$) and the antenna configuration.   
The resulting images have highest spatial resolution in the 
north-east-south-west direction.  
The respective rms noise is 0.34 and 0.36~mJy/beam.  
The signal-to-noise level reached in the images
is 1220 at the lower frequency and 880 at the higher frequency.  
The contour levels displayed in the figure are 
3, 10, 30, 100, and 300 times the rms noise.  
Additionally, the 1000$\sigma$ contour is shown for 
the 302~GHz continuum and the 600$\sigma$ contour shown for the 
346~GHz continuum.  

The total continuum flux density is 2.14 and 2.23~Jy at 302 and 346~GHz, respectively.  
Using the flux measured at $\nu = 302$~GHz and assuming that the continuum emission is optically thin, 
the total dust mass, $M_\mathrm{dust}$, can be estimated as 
\begin{equation}
M_\mathrm{dust} = \frac{D^{2} F_\nu}{\kappa_\nu B_\nu(T_\mathrm{dust})}, 
\label{dustmass}
\end{equation}
where $D$ is the source distance (160~pc), $\kappa_\nu$ is the dust mass opacity at a frequency, $\nu$, 
and $B_\nu$ the value of the Planck function for a dust temperature, $T_\mathrm{dust}$. 
For $\kappa_\nu = 5$~cm$^{2}$~g$^{-1}$ and a dust temperature of 
30~K \citep{beckwith90,andrews11}, this suggests a total dust mass of $1.3\times10^{30}$~g 
which corresponds to 0.70~$M_\mathrm{Jup}$. 
Because HD~97048 is a Herbig~Ae star, the dust temperature in the outer disk could be 
higher than 30~K; hence, this estimate is an upper limit to the total dust mass for 
the assumed dust mass opacity. 

The two fluxes measured by ALMA (2.14 and 2.23~Jy at 302 and 346~GHz, respectively) 
imply a very shallow spectral index for the continuum emission ($\alpha = 0.30\pm0.03$, 
where $F_\nu \propto \nu^{\alpha}$). 
This also suggests that the emission is highly optically thick at these frequencies, 
meaning that the primary assumption made in deriving the dust mass estimate no longer holds.   
However, given the issues with the flux calibration of the data at 346~GHz data 
detailed in Section~\ref{observations}, it is possible that we have underestimated the 
error on the flux at this frequency.  
We leave a more robust determination of the spectral index of the 
dust emission (and also the dust mass) to future work using data at higher spatial 
resolution and additional data at lower frequencies.  

\footnotetext{\url{http://casa.nrao.edu/docs/TaskRef/clean-task.html}}

\begin{figure*}
\subfigure{\includegraphics[width=0.5\textwidth]{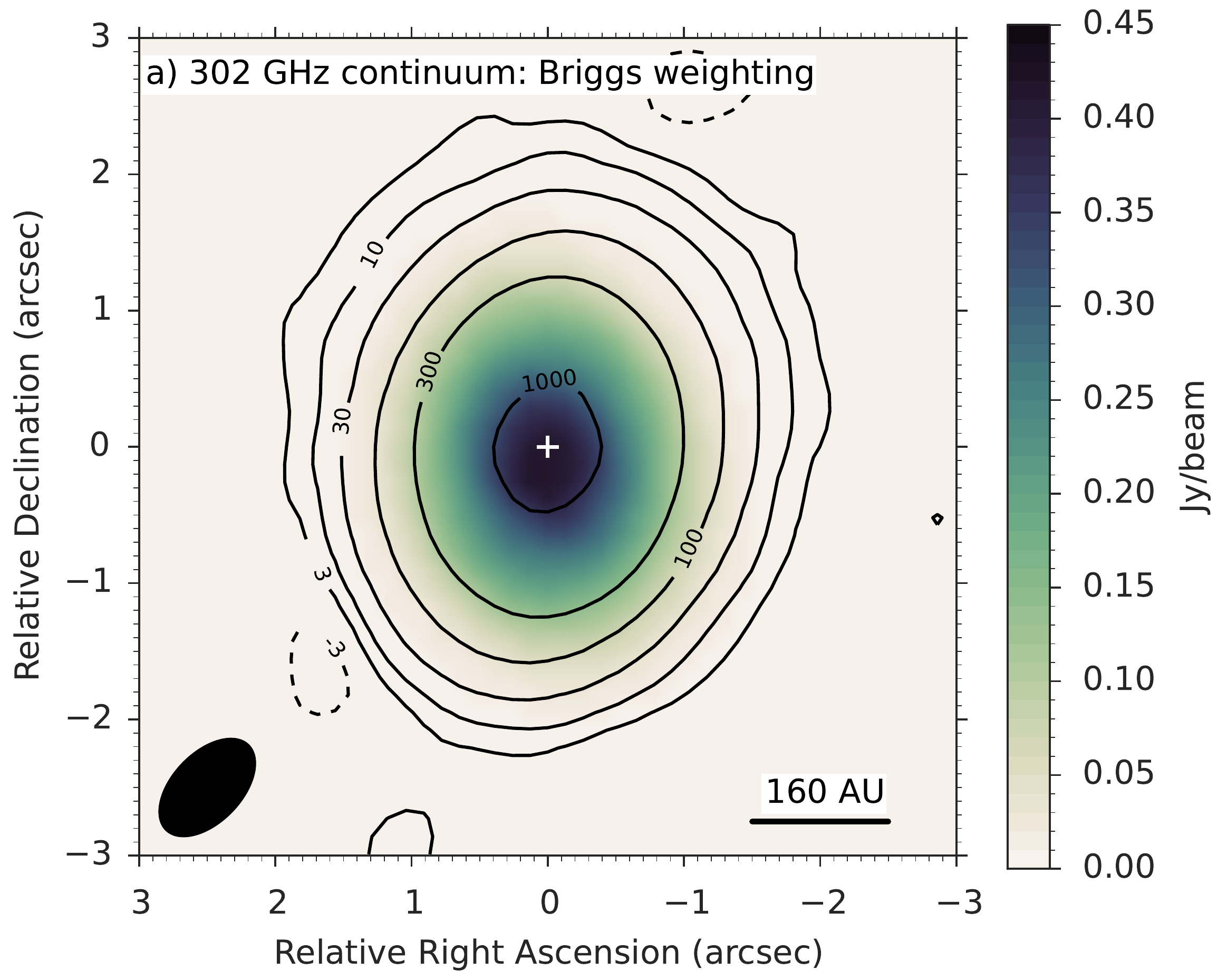}}
\subfigure{\includegraphics[width=0.5\textwidth]{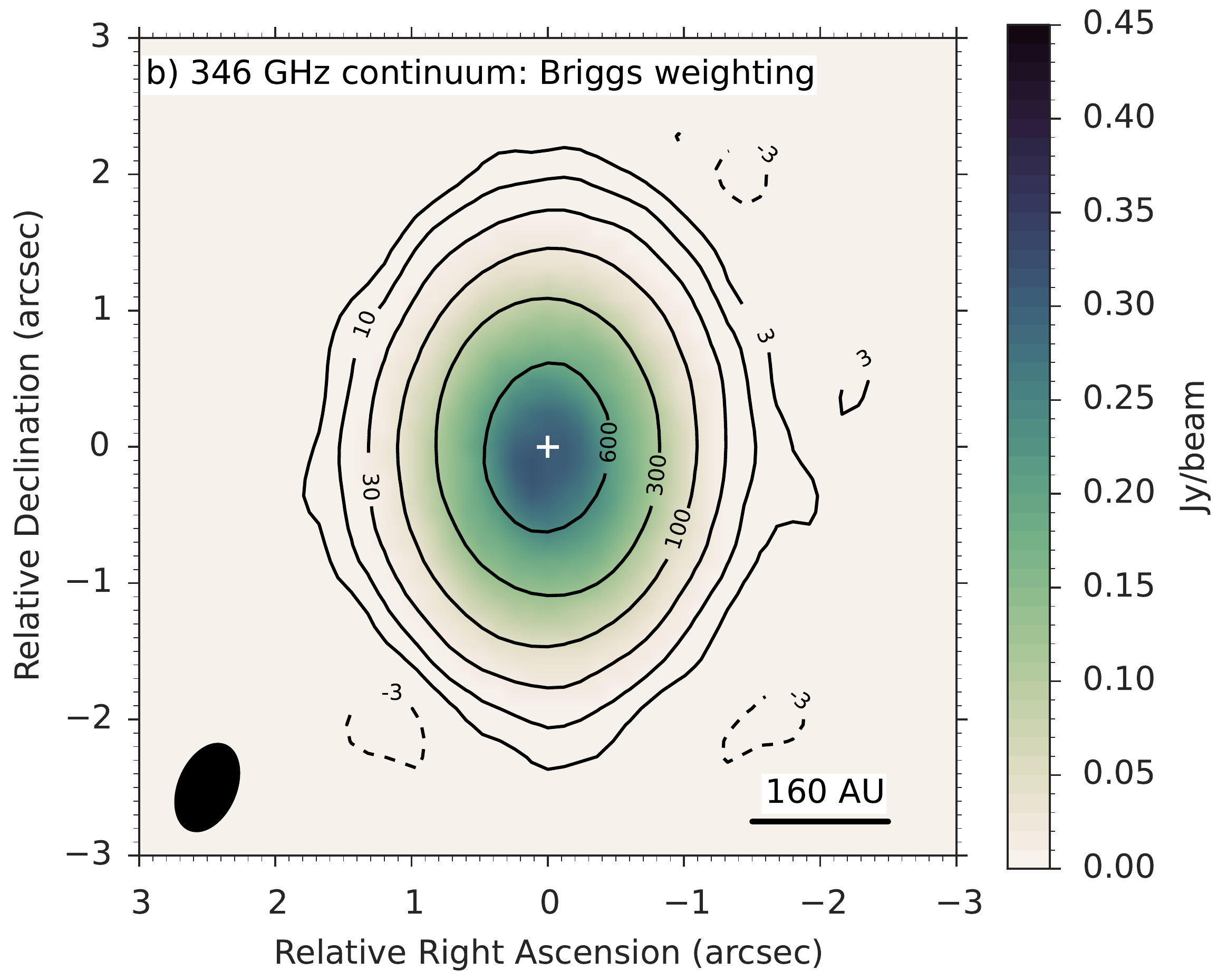}}
\subfigure{\includegraphics[width=0.5\textwidth]{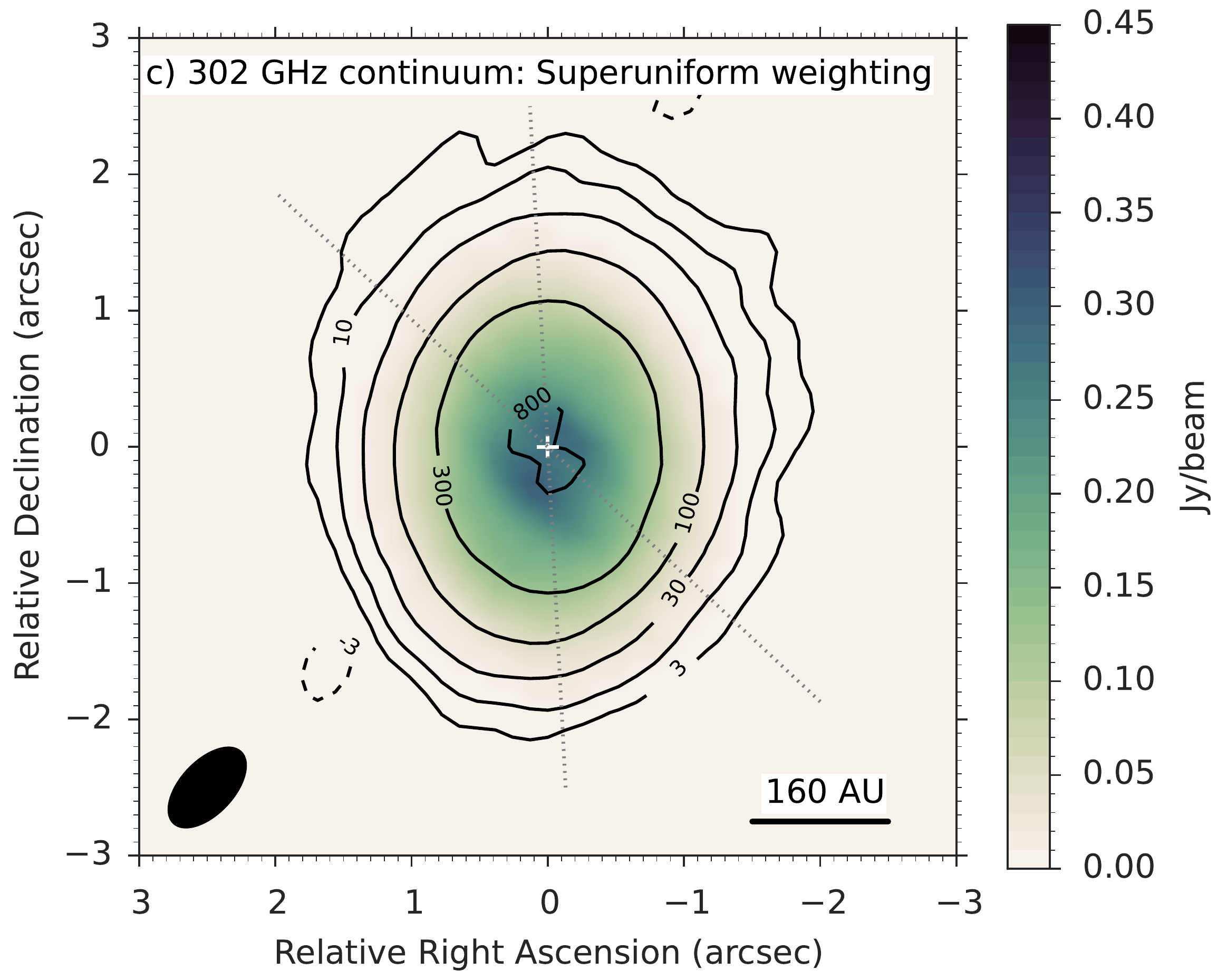}}
\subfigure{\includegraphics[width=0.5\textwidth]{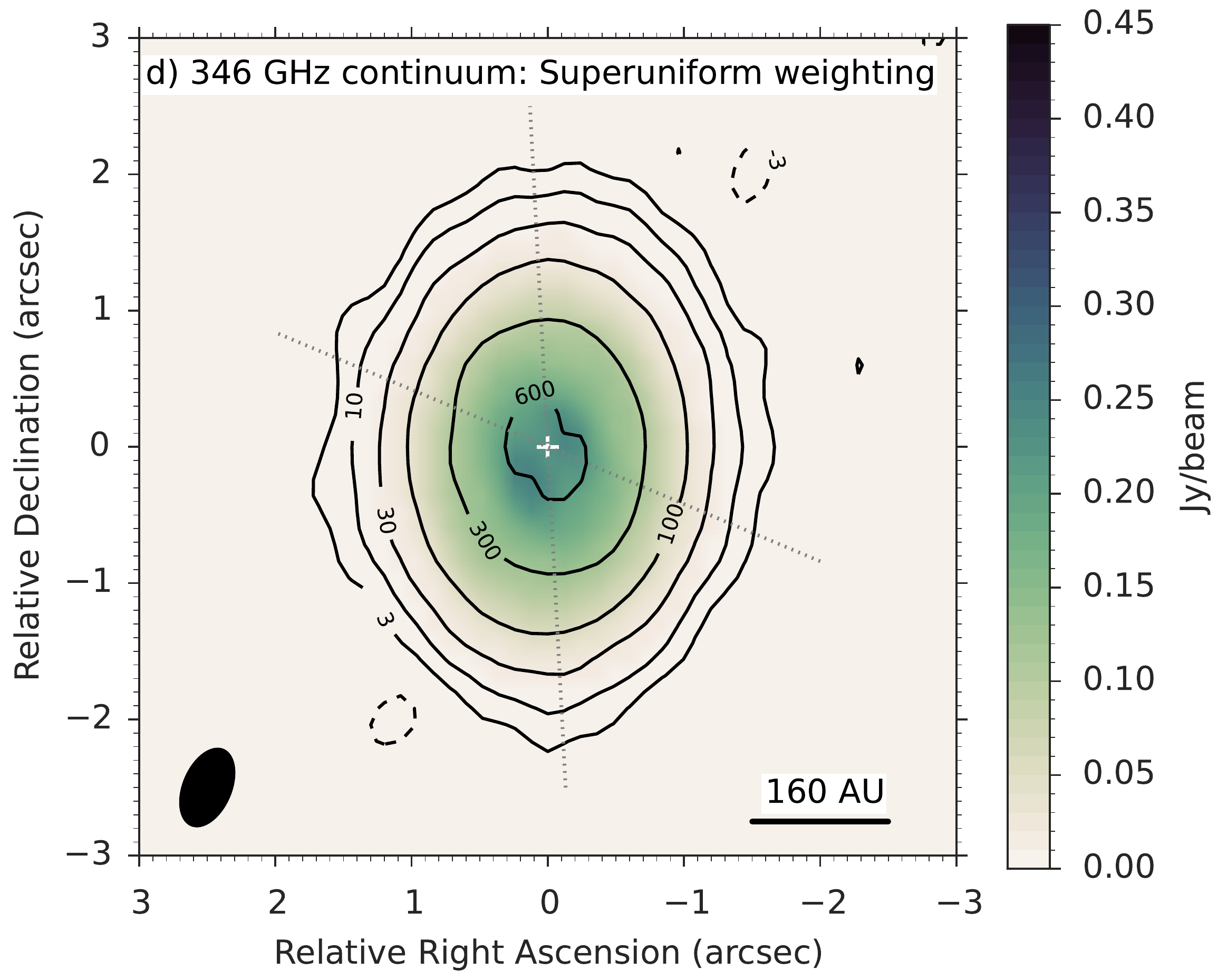}}
\caption{ALMA Cycle 0 continuum images of the disk surrounding HD~97048 at 302~GHz (left-hand panels) and 
346~GHz (right-hand panels), CLEANed using Briggs weighting (top row) and superuniform weighting 
(bottom row).  
The synthesized beam sizes are listed in the main text.
The white cross indicates the source position.  
The dotted gray lines in the bottom two panels show the 
slices across the disk major axis ($3\degree$) and the 
beam minor axes ($47\degree$ and $67\degree$ at 302 and 346~GHz, respectively). 
The contour levels are in $\sigma$, where $1\sigma= $0.34 and 0.36~mJy, respectively.}
\label{figure1}
\end{figure*}

The images in Figure~\ref{figure1} show a spatially resolved and extended dusty disk 
at least 2\arcsec~in radius, corresponding to at least 320~au, assuming 
a distance to source of 160~pc.   
As a first step in characterizing the system, the CASA task \texttt{uvmodelfit}\footnotemark, 
was used to fit the continuum visibilities assuming that the emission arises from an elliptical disk.  
The two measurement sets (at 302 and 346~GHz) were fit independently and resulted in 
almost identical disk parameters: a disk radius of 2\farcs2 (350~au), 
an inclination of $41\degree$, and a position angle (P.A.) of $3\degree$.  
The fit also confirmed that the phase center of the data is at the expected source 
position.  
The fitted P.A.~is consistent with the images: an inclined disk with a P.A.~close 
to $0\degree$ will naturally appear elongated in the north-south direction relative to 
the east-west direction.  
The images and fitting also confirm that the continuum emission is arising predominantly 
from the disk midplane because the aspect ratio is that expected for an 
ellipse centered at the source position with the disk P.A. and inclination.  
On the other hand, emission from a flared surface from such an inclined disk with 
P.A.~$\approx 0$ would show an extension in emission toward the east 
relative to the west \citep[see, e.g.,][]{lagage06}.
The disk inclination is in good agreement with previous estimates 
derived from mid-IR and PDI images \citep[][]{lagage06,quanz12}.  
Previous determinations of the P.A. of the disk 
(east of north) range from $160\degree\pm19\degree$ \citep{acke06} 
to $78\degree\pm10\degree$ \citep{quanz12}.  
These values differ somewhat from that determined here; however, 
the observations from which the P.A. is derived are 
more sensitive to the inner disk.  
As discussed in \citet{quanz12}, it is possible for both the 
inner disk geometry and the dust properties to differ from those 
in the outer disk.  
Although never quantified, the mid-IR images also suggest a position 
angle close to $0\degree$ on the sky \citep{lagage06,doucet07,marinas11}.  

In the images presented in the top panels of Figure~\ref{figure1} there 
is no indication of a gap in the disk nor any hint of additional structure.  
The maximum spatial resolution achieved in the images (set by the size of the 
minor axis of the synthesized beam) is equivalent to  83~au at 
302~GHz and 69~au at 346~GHz, assuming a distance to source of 160~pc.  
Given that the proposed size of the cavity in the inner disk from mid-IR imaging 
is $34\pm4$~au in radius \citep{maaskant13}, the image resolution is just beyond 
that required to detect an equivalent-sized gap in the (sub-)mm images. 
The use of Briggs weighting in the CLEAN algorithm is a compromise between 
sensitivity to large scale emission and spatial resolution 
\citep[natural versus uniform weighting, respectively;][]{briggs95}.
Because the signal-to-noise level of the images is high, we can apply the 
uniform weighting scheme in the CLEAN algorithm which gives higher resolution 
at the expense of a reduced signal-to-noise \citep[see, e.g.,][]{isella12}.  
In the bottom panels of Figure~\ref{figure1}, we present the resulting images 
using superuniform weighting.  
This has the added effect that the side lobes are further suppressed, 
allowing better sensitivity on the largest scales compared with uniform weighting 
\citep{briggs95}.  
The resulting synthesized beam sizes are 0\farcs73$\times$0\farcs40 (-43\fdg3)
and 0\farcs61$\times$0\farcs36 (-23\fdg0) at 302 and 346~GHz respectively.  
The spatial resolution is increased to 
64~au for the 302~GHz continuum and 58~au for the 346~GHz continuum.  
The images now show the signature of a gap (or at least a reduction) in the emission 
on small scales indicated by the shapes of the 
$800\sigma$ and $600\sigma$ contours at 302 and 346~GHz, respectively.  
The ``hour-glass''-shaped innermost contours, which are aligned with the beam minor axis along 
which the spatial resolution is highest, 
show that the peak of the emission is offset from the 
source position \citep[indicated by the white cross; see also][]{hughes07,andrews11}.  

\footnotetext{\url{http://casa.nrao.edu/docs/taskref/uvmodelfit-task.html}}

In Figure~\ref{figure2}, we display slices through the continuum 
images generated using superuniform weighting along the disk major 
axis ($3\degree$, dark blue lines) and along the beam minor 
axes at each frequency (light green lines).  
For the latter data, the offset has been scaled to account for 
the disk inclination ($41\degree$) and the respective position angle of the beams.\footnotemark  
~A positive offset indicates the direction of increasing decl. and 
R.A. 
The slices are indicated by the dashed gray lines in the bottom panels 
of Figure~\ref{figure1}.  
All slices show a flattening and/or reduction in flux at the position of the 
source, again showing that the continuum emission peaks at a position  
offset from the source center.  

\begin{figure*}
\centering
\includegraphics[width=\textwidth]{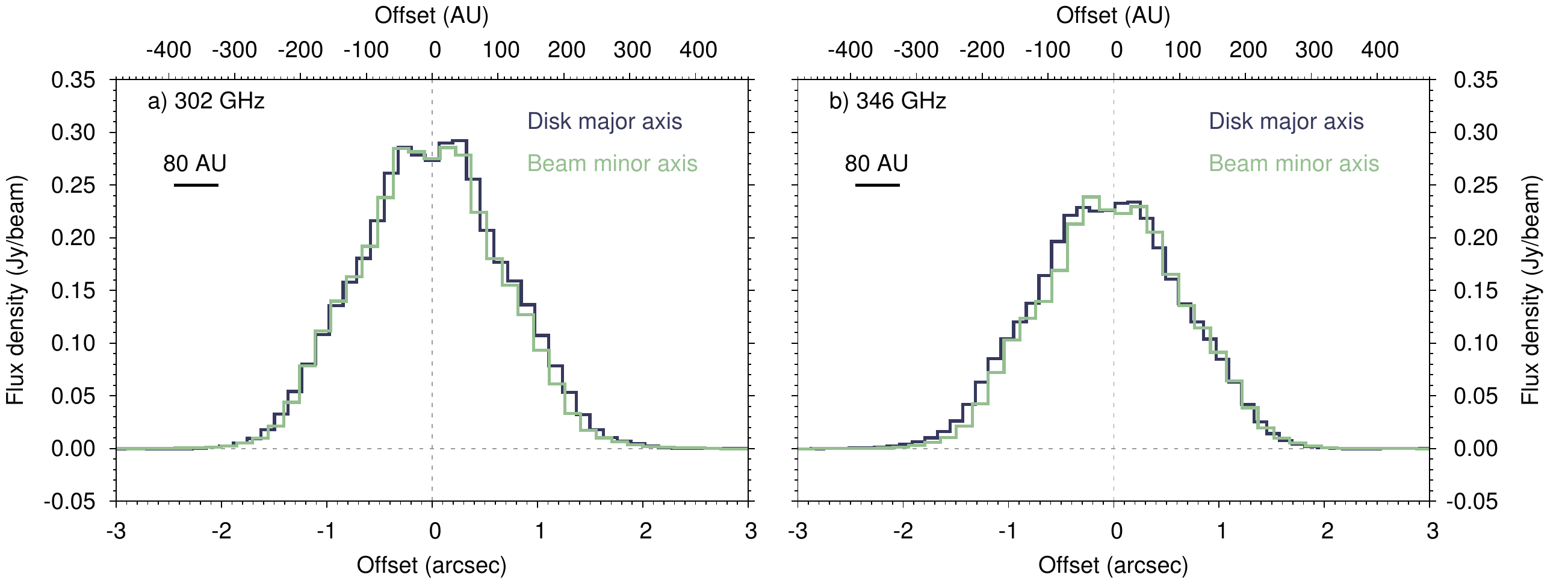}
\caption{Continuum flux density at 302~GHz (left-hand panel) and 346~GHz (right-hand panel) 
along a slice across the disk major axis ($3\degree$, dark blue lines) and beam minor axes 
($47\degree$ and $67\degree$ at 302 and 346~GHz, respectively, light green lines). }
\label{figure2}
\end{figure*}

\footnotetext{$(r/a) = \cos i/\sqrt{\sin^2 \alpha + \cos^2\alpha \cos^2 i}$ where $r$ is the projected 
radius of the disk along the minor axis of the beam, $a$ is the disk major axis 
(assumed aligned in the north-south direction), $i$ is the disk inclination, and $\alpha$ is the angle of the 
minor axis of the beam, measured east from north.}

\subsubsection{Continuum visibilities}
\label{visibilities}

We conduct all subsequent analysis of the continuum emission in the 
visibility domain.  
This allows extraction of information about the disk structure 
on spatial scales smoothed over in the reconstructed images.  
Whether the continuum emission originates from a disk with a dust-depleted inner cavity (i.e. a ring), 
as suggested in the images with superuniform weighting, can be tested by examining the visibilities.  
For a circular-symmetric brightness distribution from a ring, the real components of 
the visibilities will show a null (zero crossing) and the 
imaginary components will be close to zero \citep[see, e.g.,][]{hughes07,walsh14}.  
In Figure~\ref{figure3}, we display the 
visibilities for HD~97048 which have been rotated and deprojected 
using the disk inclination and P.A. determined from the ALMA data 
($41\degree$ and $3\degree$, respectively).  
The top panels show the real components and the bottom panels show the 
imaginary components, both binned into 10~k$\lambda$-sized bins.  
The error bars shown correspond to the standard error of the mean and are 
generally smaller than the point size.  
The real components show a shallow null at $\approx150$~k$\lambda$ indicating 
that the emission may arise from a ring.  
This is better seen in the inset plots which zoom into the region 
specified by the dashed gray box in the main plot.  
The imaginary components show small scatter about zero, confirming that the continuum 
emission is close to circular symmetric.  
On the longest baselines ($\gtrsim250$~k$\lambda$), the imaginary components begin 
to deviate from zero. 
This is most apparent in the higher signal-to-noise data at 302~GHz.  
However, the magnitude of the largest imaginary components is on the order of 
20\% of the real components; hence, for the most part, the emission arises 
from a circular symmetric disk.    

\begin{figure*}
\centering
\includegraphics[width=\textwidth]{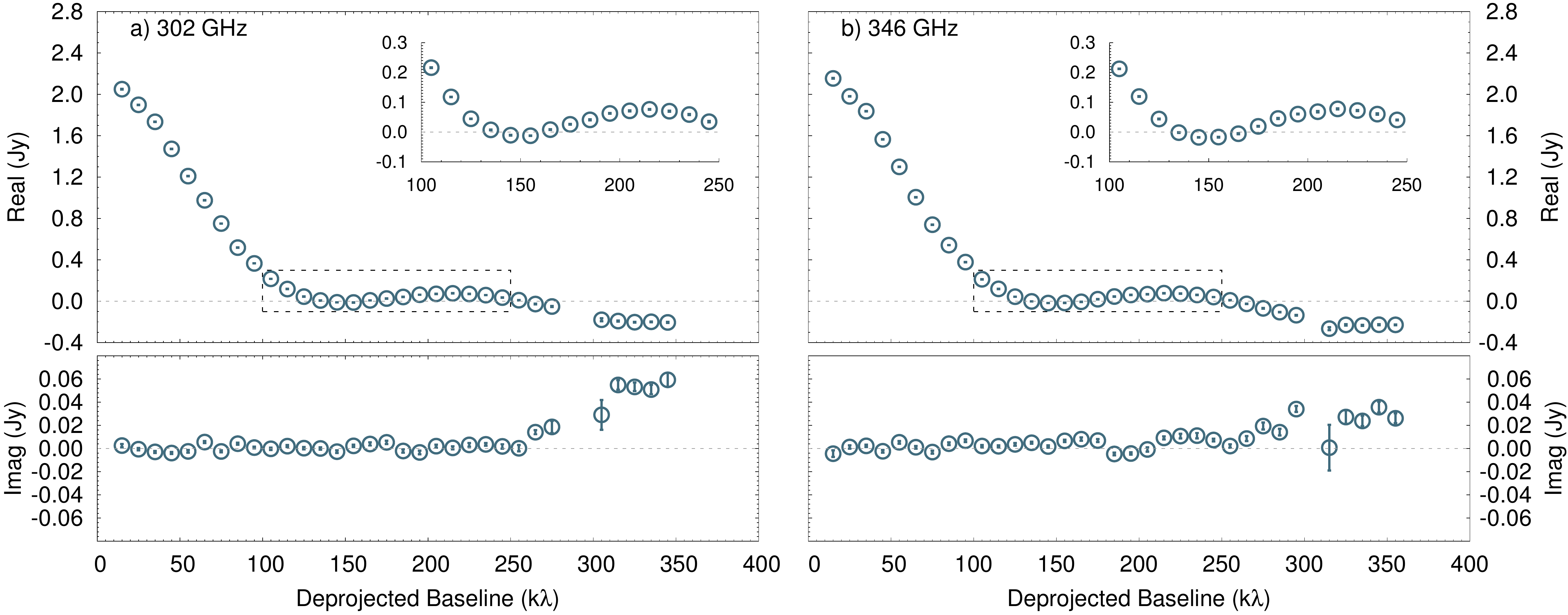}
\caption{Real and imaginary components of the visibilities (top and bottom, respectively) 
at 302~GHz (left-hand panels) and 346 GHz (right-hand panels).  
The data are binned to 10 k$\lambda$-sized bins and the error bars correspond to the 
standard error of the mean in each bin.  
Note that the error bars are, in general, smaller than the size of the points. 
The inset plots show a zoom of the region enclosed by the dashed gray box.}
\label{figure3}
\end{figure*}

\subsubsection{Continuum modeling}
\label{modelling}

To quantify the morphology of the disk we model the 
brightness distribution using simple analytical functions.  
The real components of the visibilities are then calculated given that 
the intensity distribution can be treated as a summation over 
infinitesimally narrow rings, the Fourier transform for which is a 
zeroth-order Bessel function of the first kind \citep[$J_0$,][]{berger07}.  
The real components are thus given by,
\begin{equation}
V_\mathrm{Re}(\rho) = 2\pi \int_0^{\infty} I(r) \, J_0(2\pi\,r\,\rho) \, r \, dr,
\label{equation1}
\end{equation}
where $\rho$ is the baseline grid (for which the binned data grid was used).  
Any circular-symmetric intensity distribution, $I(r)$, can be inserted into this 
equation to compute the corresponding real components of the visibilities.  
Note that we do not attempt to decouple the dust surface density and 
temperature profiles because to do so would introduce significant degeneracies 
in the models.  
To break such degeneracies requires spatially resolved 
continuum observations at two well-separated frequencies and even then, 
assumptions have to be made regarding the dust opacity 
\citep[see, e.g.,][]{andrews07,guilloteau11}.  
We leave characterization of the dust density and temperature distribution 
to future work at higher resolutions.  

To extract the intensity profiles at both frequencies from the binned and deprojected 
visibilities, we use the method recently proposed by \citet{zhang16}.  
This involves modeling the intensity distribution as a summation over 
Gaussians (each with an amplitude, $a_i$, and a width, $\sigma_i$) modulated 
by a sinusoidal function with a spatial frequency, $\rho_i$.  
The underlying intensity distribution is assumed to also be a
gaussian function with a width, $\sigma_0$.  
This technique is powerful because it allows the addition of 
substructure (including peaks and troughs) with various spatial frequencies, 
whilst also ensuring that the intensity tends to 0 at infinity.  
Using this method, \citet{zhang16} were able to recover the intensity profile in 
four protoplanetary disks recently imaged at moderate 
($\approx$ 0\farcs2 -- 0\farcs4) to high ($\approx$ 0\farcs02 -- 0\farcs03) 
angular resolution. 
A particular highlight is that they recover the major features of the 
intensity profile of the dust emission from HL Tau, 
using only baselines $\lesssim 2000$~k$\lambda$.  
The produced image, on the other hand, was created using 
baselines $\lesssim 12,000$~k$\lambda$ \citep{alma15}.
Further details on the methodology can be found in \citet{zhang16}.  

The radial intensity profile is given by
\begin{equation}
I_{n}(r) = \frac{a_0}{\sqrt{2\pi}\sigma_0} \exp{\left(-\frac{r^2}{2\sigma_0^2}\right)} \, + \, 
 \sum_{i=1}^{n}\cos{(2\pi r \rho_i)} \times \frac{a_i}{\sqrt{2\pi}\sigma_i} \exp{\left(-\frac{r^2}{2\sigma_i^2}\right)}.   
\end{equation}
\citet{zhang16} use the Levenburg-Marquardt $\chi^2$ minimization technique 
to fit their intensity profiles. 
Because this technique is prone to getting stuck in local minima unless 
a good starting point is found, we opt to use the Markov Chain Monte 
Carlo (mcmc) approach coupled with Bayesian statistics, 
to determine the best-fit intensity profile, $I_n(r)$, 
for a given number of gaussian components, $n$.  
A radial grid of 10~au resolution was deemed sufficient to 
model the data.  
The Python module, pymc\footnotemark, was used \citep{patil10}.  
We assume all prior distributions for the fitted 
parameters, $[a_i, \sigma_i, \rho_i]$, 
are uniform with boundaries set by the range of 
radii and spatial frequencies probed by the continuum data 
($\lesssim 500$~au and $\lesssim 500$~k$\lambda$, respectively).  
The intensity profiles are also scaled to the observations 
using the observed flux at both frequencies, which 
are assumed to have a normal distribution with a standard deviation 
of 10\% at 302~GHz and 15\% at 346~GHz.  

Our modeling approach is now described and is similar to 
that conducted by \citet{zhang16}.  
First, a single gaussian intensity profile is fitted, $I_0$, 
to determine the best-fit gaussian width, $\sigma_0$, 
and to test whether this is an adequate fit to the data.  
Additional single gaussian components are then systematically added  
to the fit, using the best-fit parameters from the previous 
model (because MCMC models also benefit from a good starting point).   
The addition of components is halted when an incremental 
improvement in the real residuals of the visibilities is found.  
We also go one step further than \citet{zhang16} and display the 
residuals from the model fitting in both the real components of 
the visibilities, and the corresponding residual images. 

It is found that two additional Gaussian components only are required 
(i.e., $I_2$) to reproduce the visibility 
data at both frequencies. 
Figure~\ref{figure4} shows the resulting visibility profiles at both 
frequencies and respective real residuals.  
The model fits give an excellent ``by-eye" fit to the visibility data: 
quantitatively, the fits are also good, with peak residuals on the 
order of $\approx20-30$~mJy (corresponding to $\approx 1$\% of the total 
flux density).  
Figure~\ref{figure5} shows the imaged residuals at 302 and 345~GHz.  
The peak residuals are 16 and 23~mJy/beam, respectively, which 
correspond to 4\% and 7\% of the peak flux density in the observed 
images.  
The maximum imaged residuals are also close to, and offset from, 
the source position.  
This could be due to emission arising from the innermost regions which is 
not well described using a circular-symmetric intensity profile.  
That the imaginary components deviate from zero on the longest baselines 
(also visible in the unbinned data)
is also possible evidence of non-axisymmetric emission on small scales 
(see Figure~\ref{figure3}).  
Furthermore, the total flux densities in the imaged residuals are 8.0 and 26~mJy/beam, 
so that the models reproduce $\approx 99$\% of the observed fluxes 
(2.14 and 2.23 Jy, respectively).  

In Figure~\ref{figure6} we show the extracted and normalized intensity profile 
for the 302~GHz data at the model grid resolution (10~au, light green solid lines), 
and convolved with a 20~au beam (dotted gray lines) and a 
50~au beam (dashed dark blue lines) on both a linear (left) 
and a logarithmic (right) intensity scale.  
The dashed horizontal line on the logarithmic plot represents 
the normalized $3\sigma$ rms continuum noise at 302~GHz.  
The respective plots for the 346~GHz data fits are very similar and so are 
not displayed.  
The fits to the visibility profile predict the presence of 
two peaks in the radial intensity profile at 
$\approx50$ and 150~au, and a third 
``shoulder" at $\approx250-300$~au is evident in the logarithmic plot.  
The model fits also suggest a decrement in (sub-)mm continuum emission 
within $\approx50$~au, confirming the classification of this disk as 
a transition disk, as well as gaps at $\approx100$~au and $250$~au.  
We note that we tried a more complex model with an additional 
central component to attempt to quantify the decrement in emission 
within the cavity; however, the improvement in fit was not significant, 
and higher spatial resolution data are needed to further guide the models.
We emphasize here that the recovery of this substructure is only 
possible due to the high signal-to-noise of the data ($\gtrsim1000$) 
and the direct modeling of the visibility data.

\footnotetext{\url{https://pypi.python.org/pypi/seaborn/0.6.0}}

\begin{figure*}[!h]
\includegraphics[width=\textwidth]{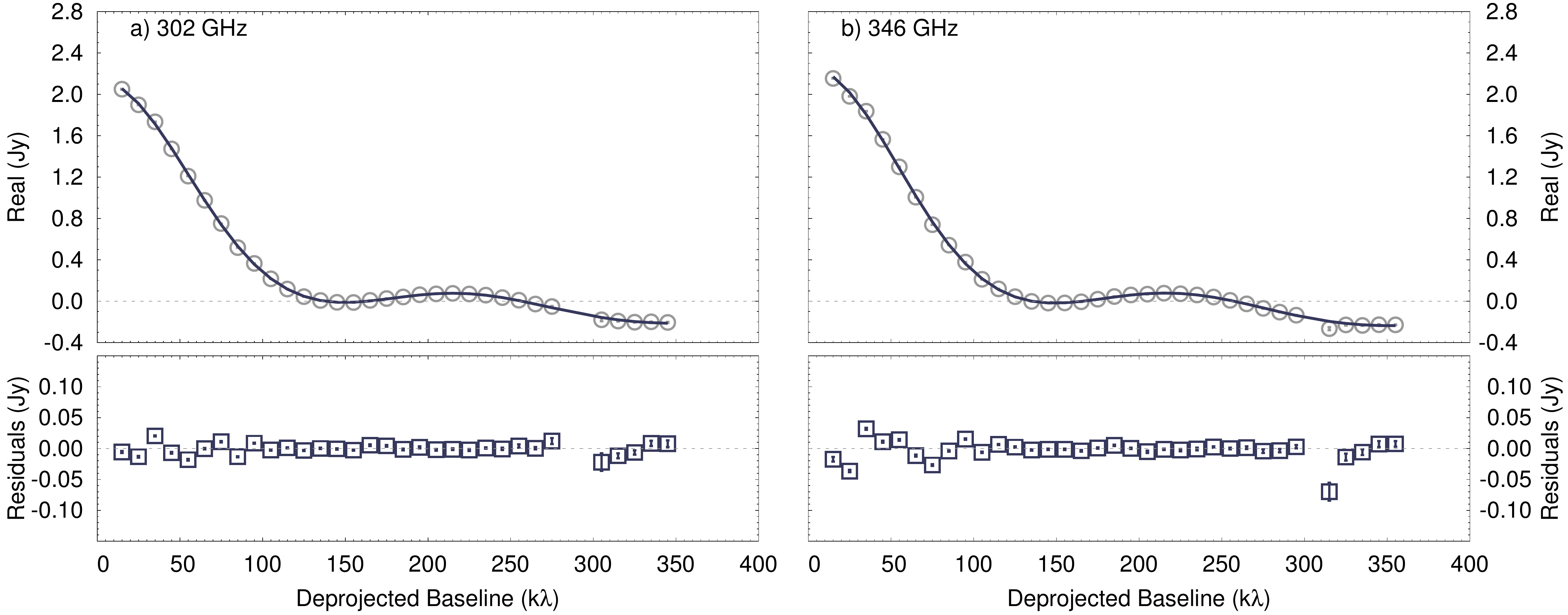}
\caption{Real components of the visibilities for the best-fit intensity profiles 
(dark blue lines) and respective real residuals as a function of deprojected baseline.  
The observed binned and deprojected visibilities are plotted in gray circles.}
\label{figure4}
\end{figure*}

\begin{figure*}
\centering
\subfigure{\includegraphics[width=0.45\textwidth]{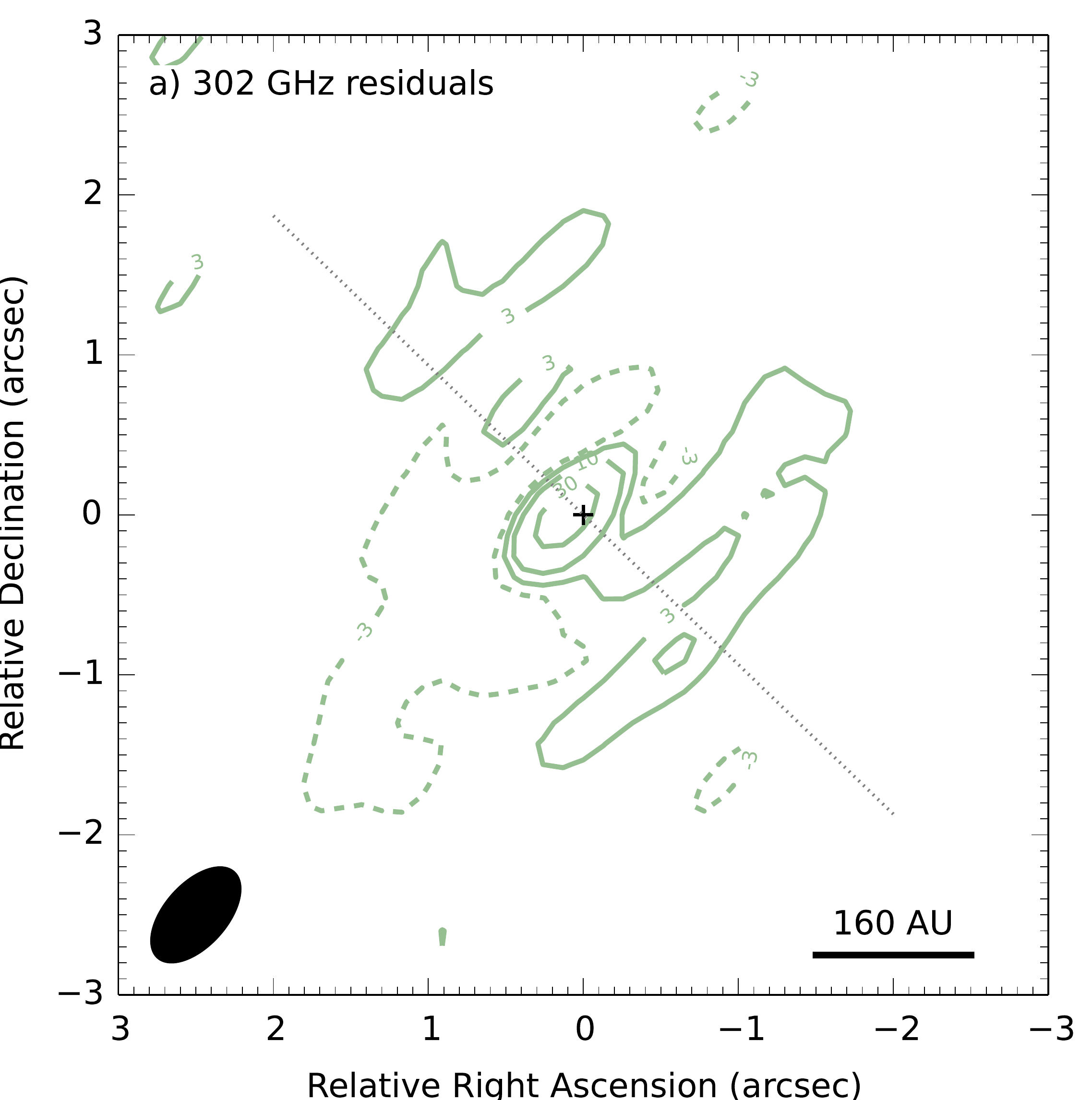}}
\subfigure{\includegraphics[width=0.45\textwidth]{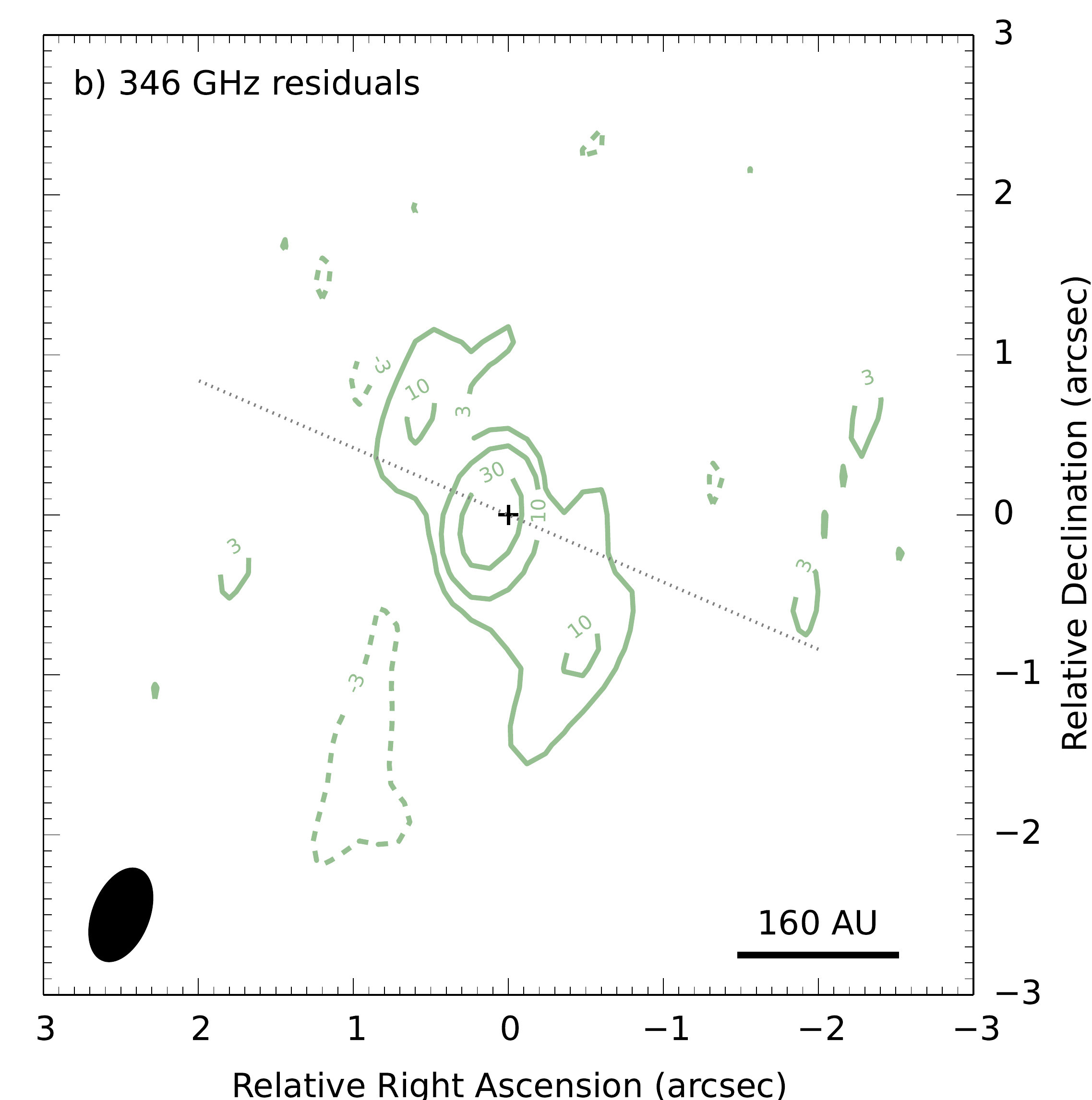}}
\caption{Real residuals imaged with superuniform weighting at 302~GHz (left-hand panels) and 
at 346~GHz (right-hand panels) for the best-fit intensity profiles.  
The black cross indicates the source (or stellar) position.  
The dotted gray lines show the slices across the beam minor axes 
($47\degree$ and $67\degree$ at 302 and 346~GHz, respectively). 
The contour levels are in $\sigma$, where $1\sigma=$ 0.34 and 0.36~mJy, respectively.}
\label{figure5}
\end{figure*}

\begin{figure*}[]
\centering
\includegraphics[width=\textwidth]{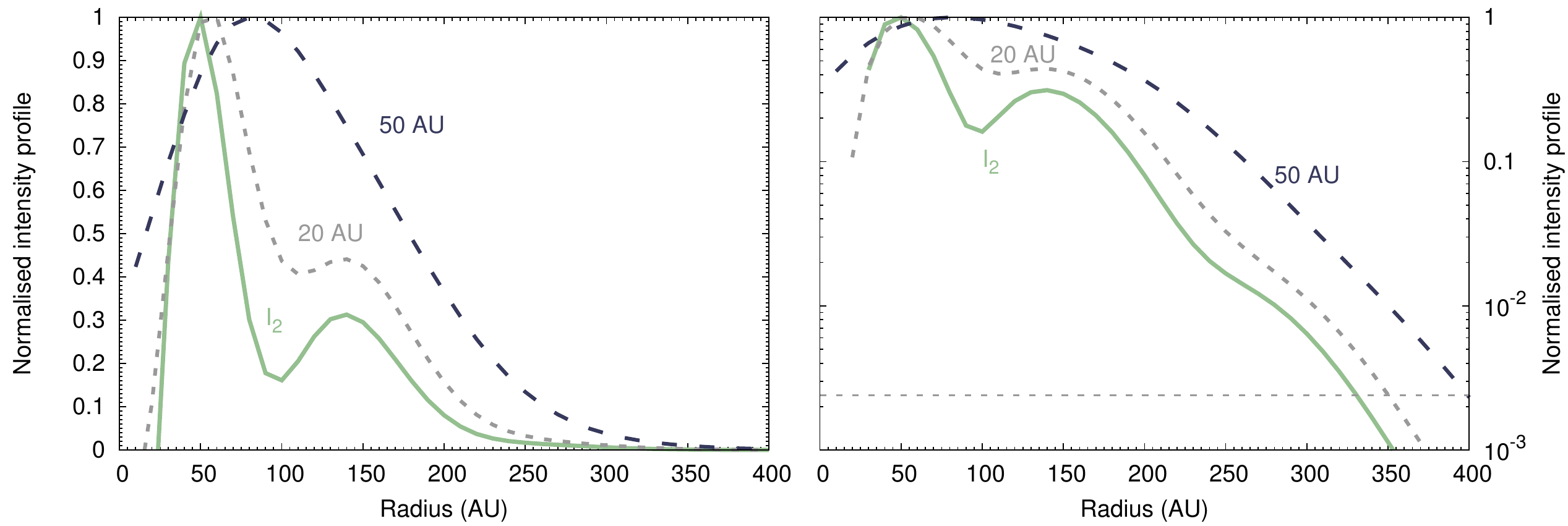}
\caption{Normalized continuum intensity profiles as a function of radius for the best-fit 
intensity profile at the model grid resolution (light green solid lines) and convolved with a 
20~au beam (gray dotted lines) and a 50~au beam (dark blue dashed lines).  
The horizontal dashed gray line in the right-hand panel signifies the $3\sigma$ rms continuum 
noise at 302~GHz (relative to the peak intensity).}
\label{figure6}
\end{figure*}

\subsection{CO $J=3$-2 line emission}
\label{coline}

\subsubsection{Channel maps and line profile}

The CO $J=3$-2 (345.796~GHz) emission was extracted from the full 
self-calibrated measurement set and then continuum subtracted 
in the $uv$-plane using the CASA task \texttt{uvcontsub}.\footnotemark  
~The CO emission was then imaged using CLEAN with Briggs 
weighting (robust~=~0.5), with a slight oversampling in velocity 
resolution (0.15~km~s$^{-1}$).  
The resulting synthesized beam is 0\farcs69$\times$0\farcs44 (-23\fdg2).  
Masking was done interactively on a channel-per-channel 
basis. 
Emission was detected across $\approx100$100 velocity channels ranging from 
-2.70 to 12.3~km~s$^{-1}$.  
The resulting rms noise is 18.1 mJy/beam in the channels, and the peak 
signal-to-noise is 122.

\footnotetext{http://casa.nrao.edu/docs/taskref/uvcontsub-task.html}

In Figure~\ref{figure8}, the CO $J=3$-2 (345.796~GHz) channel maps 
(shifted by the source velocity) are displayed.  
The maps show a spatially resolved and inclined and flared disk, 
with blue-shifted emission from the south side of the disk and 
red-shifted emission from the north side.  
The maps confirm that the disk is inclined toward the east relative 
to the disk major axis (P.A. 3\degree), and is rotating in a 
counter-clockwise direction.  
This is in agreement with the disk rotation direction determined via 
analysis of [OI] line emission at $6300\AA$~\citep{acke06}.  
Line emission is suppressed or missing in velocity channels from 
-1.20~to~0.90~km~s$^{-1}$ inclusive.   
This could be due to absorption by foreground material 
\citep[the observed extinction toward HD~97048 is 1.24~mag,][]{vandenancker98}, 
or spatial filtering of extended emission from background material.    
HD~97048 is in close proximity to the Chameleon~I molecular cloud. 

\begin{figure*}
\includegraphics[width=\textwidth]{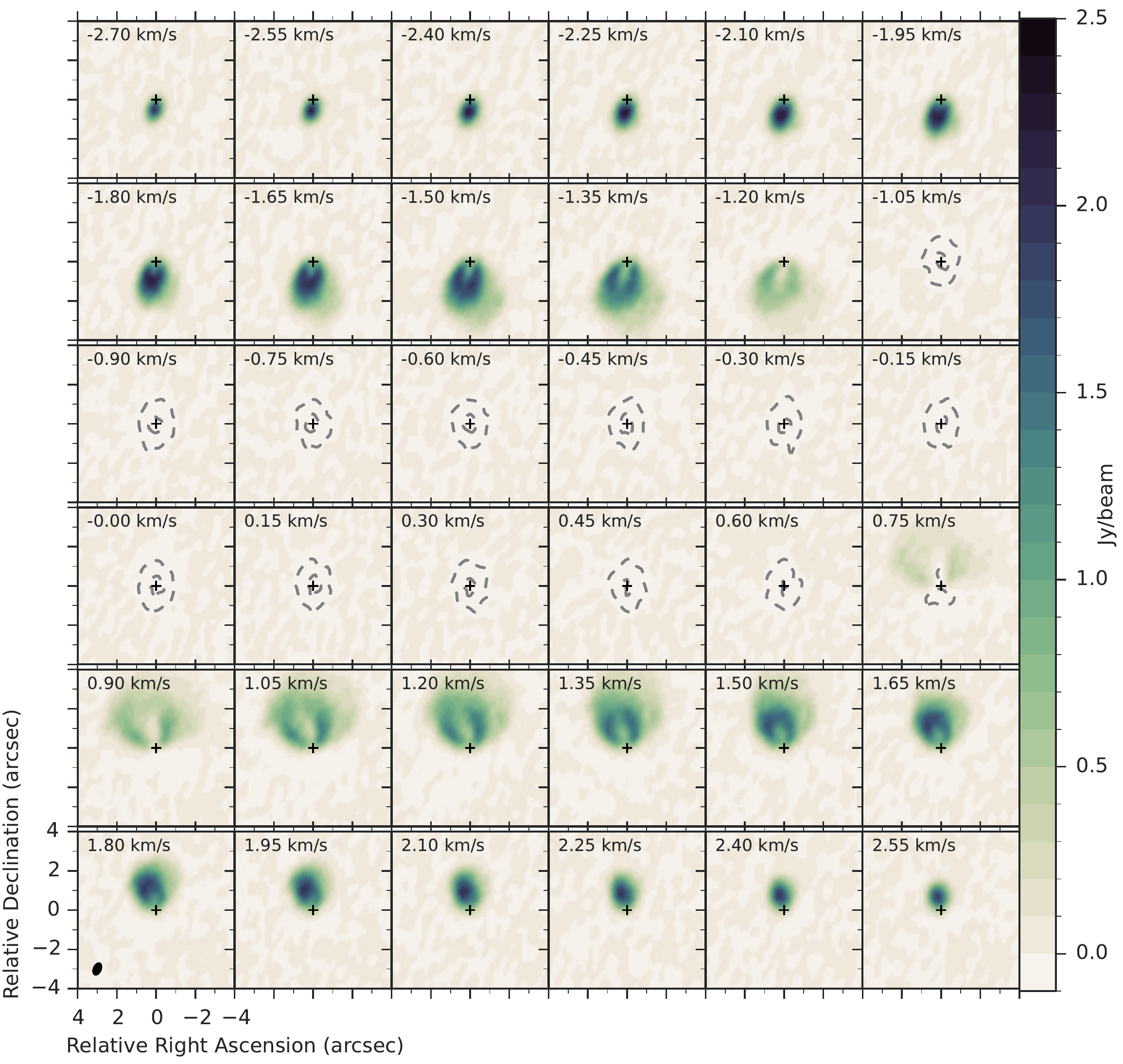}
\caption{Channel maps showing the CO $J=3$-2 (345.796~GHz) line emission.  
The missing emission from -1.20 to 0.90 km~s$^{-1}$ inclusive may be due 
to absorption by foreground material and/or spatial filtering of extended 
background material.  
The black cross indicates the source position and the synthesized beam 
is shown in the bottom left-hand corner of the bottom left-hand panel.  
Note that the source velocity (4.65 km~s$^{-1}$) has been subtracted from 
these data.}
\label{figure8}
\end{figure*}

Even though HD~97048 lies in a region of bright extended emission, 
these data are sufficiently high quality 
to constrain the source velocity ($V_\mathrm{LSRK} = 4.65\pm0.075$~km~s$^{-1}$, 
relative to the Kinematic Local Standard of Rest).    
In Figure~\ref{figure9} we show the line profile generated 
summing all emission within a polygon, the shape of which is 
determined by the extent of the emission in all directions out to 
the 3$\sigma$ contour of the integrated intensity.  
The shaded region corresponds to those velocity channels affected 
by foreground absorption and/or spatial filtering.  
The line profile has been mirrored and shifted (gray lines) to confirm the 
determination of the source velocity from these data (indicated by the central 
vertical gray dashed line), despite the missing emission at, and around, 
the source systemic velocity.  

\begin{figure}[]
\centering
\includegraphics[width=0.5\textwidth]{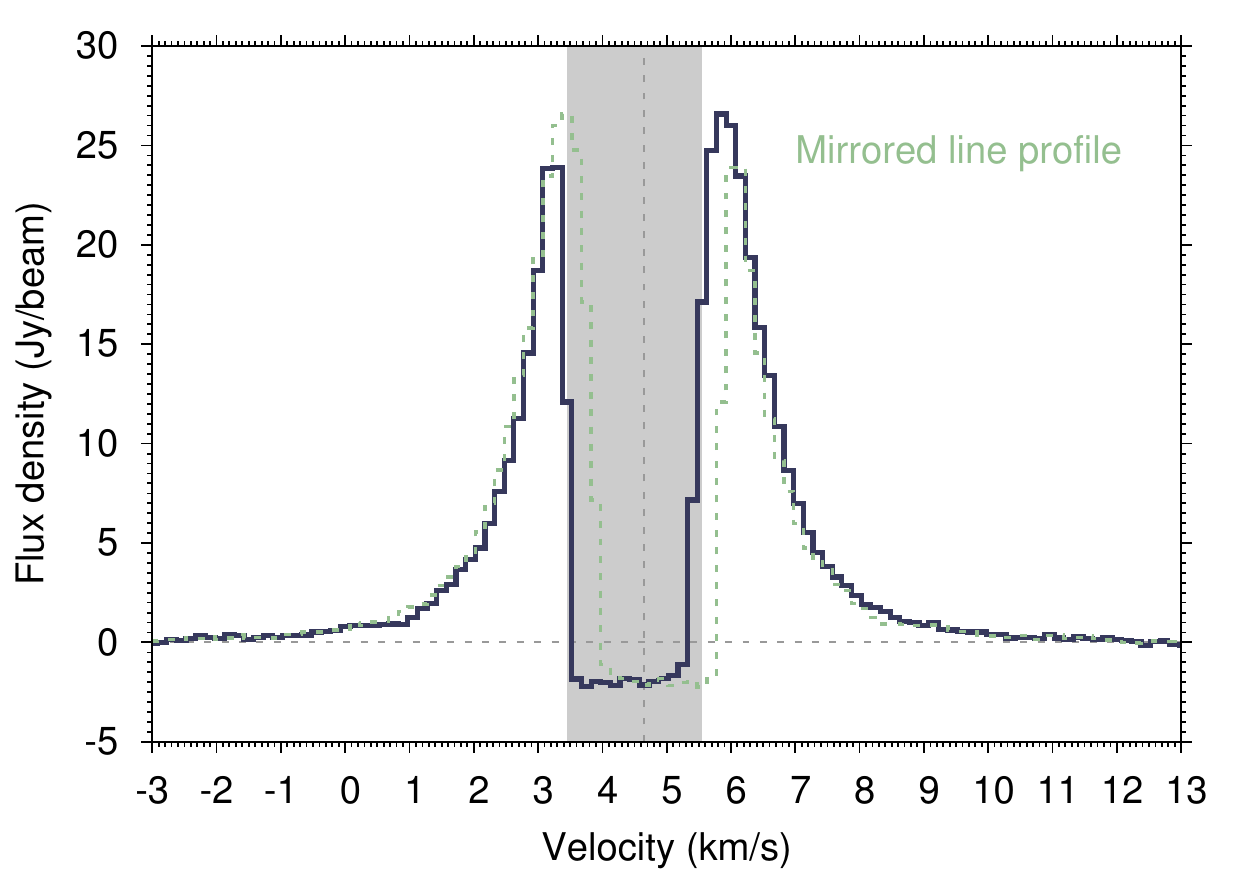}
\caption{CO $J=3$-2 (345.796~GHz) line profile generated from a polygon containing all 
emission within the maximum extent of the $3\sigma$ contour.  
The mirrored line profile is given by the dotted light greeni lines. 
The source velocity ($V_\mathrm{LSRK}$) determined from this data is $4.65\pm0.075$~km~s$^{-1}$.}
\label{figure9}
\end{figure}

\subsubsection{The radius of the molecular disk}
\label{diskradius}

In Figure~\ref{figure10}, we compare the CO $J=3$-2 first moment map 
(corrected by the source velocity)
with the integrated intensity (dashed gray contours at 3, 10, 30, 100, and $150\sigma$) 
and the $3\sigma$ continuum extent (dashed black contour).  
The first moment map was generated using a conservative clip at a $5\sigma$ level, 
whereas the integrated intensity (or zeroth moment map) 
was generated by summing over all channels in the data cube with all 
pixels with a value $<3\sigma$ masked.  
We estimate the rms noise level in the integrated intensity as 25.3~mJy~km~s$^{-1}$  
calculated by summing over {\em all} channels (87 in total) containing emission 
at a level $\ge 3\sigma$ ($\sqrt{n}\sigma\Delta v$, where 
$n$ is the number of channels, $\sigma$ is the rms noise per channel, 
and $\Delta v$ is the channel width). 
It is this value that is used to compute the dashed gray contours in 
Figure~\ref{figure10}.

The strange ``dumbbell" shape in the integrated intensity and first moment maps 
is because of the missing emission at and around the source velocity.  
The emission along the minor 
axis of the disk (east-west direction) is most 
affected, whereas that along the major axis (north-south direction) 
is least affected because it is in this direction that the projected 
velocity along the line of sight is largest.  
There is also evidence that emission from the largest spatial scales may also be 
affected because the extent of the northern red-shifted emission is larger than that 
of the southern blue-shifted emission which is consistent with the range of 
velocities most affected by foreground absorption and/or spatial filtering (-1.20 to 0.90~km~s$^{-1}$, 
see Figures~\ref{figure8} and \ref{figure9}).  
Hence, in all subsequent analysis, it is only possible to determine a 
lower limit to the spatial extent of the molecular disk.  
Line emission from relatively abundant and optically thin isotopologues 
(e.g., $^{13}$CO or C$^{18}$O), and/or higher energy rotational transitions of 
$^{12}$CO, may help to confirm the radial extent of the 
molecular disk. 

Figure~\ref{figure10} shows that the molecular disk extends to at least 
$\approx4''$.  
This is more extended than the continuum at 346~GHz ($\approx2''$).  
In Figure~\ref{figure11}, we compare the CO integrated intensity and 
continuum emission at 346~GHz, both normalised by their peak value, along 
the disk major axis.  
The horizontal dashed lines show the respective $3\sigma$ rms noise levels.  
This plot clearly shows that the disk is significantly more extended 
in CO line emission.  
Similar to the initial analysis conducted for the continuum, we use the 
CASA task \texttt{uvmodelfit}, to model the integrated intensity of the 
line emission as arising from a geometrically thin disk to refine the 
estimation of the disk radius.  
This gives a result for the radius of $>$4\farcs7 corresponding 
to $>750$~au, making this disk one of the largest known in $^{12}$CO molecular 
emission.  
The disk extent imaged in scattered light at optical wavelengths suggests an outer radius 
of $\approx640$~au assuming a source distance of 160~pc \citep{doering07}.  
Hence, it is likely that the ALMA data are probing the full spatial 
extent of the molecular disk despite the missing emission at and around the 
source velocity. 
These data also allow us to confirm that the extended emission seen in scattered 
light does indeed arise from a circumstellar disk, as opposed to a remnant circumstellar 
envelope as proposed as an alternative explanation by \citet{doering07}.  

The $^{12}$CO integrated intensity suggests that the molecular 
disk is at least twice as large in radial extent as the (sub-)mm dust disk which 
is possible evidence of radial drift of large ($\sim$~mm-sized) dust grains; 
however, extraction of the (sub-)mm dust and CO radial column densities 
are needed to confirm that the gas-to-dust mass ratio does indeed increase 
in the outermost regions of the disk. 
The alternative explanation is that emission external to $\approx350$~au from (sub-)mm dust 
simply falls below the sensitivity limits of the current data.   
Extraction of CO column densities is difficult to do using $^{12}$CO alone 
because the emission is optically thick and arises from the uppermost layers in the disk, and  
observations of optically thin emission from isotopologues are necessary.
Furthermore, extraction of column densities requires a well-constrained 
radial and vertical temperature structure. 
Due to the lack of spatially resolved data, especially at (sub-)mm wavelengths, 
this does not yet exist for the disk around HD~97048. 

Alternatively, the continuum rms noise level in the outermost regions 
can be used to estimate an upper limit to the unseen dust mass, 
and thus to the dust mass surface density.  
The dust mass, $M_\mathrm{dust}^\mathrm{out}$ can be estimated using 
Equation~\ref{dustmass}: 
at $\nu=302$~GHz, the rms noise is 0.34~mJy~beam$^{-1}$ which gives
an annulus-integrated continuum flux density of $F_\nu^\mathrm{out} \approx40$~mJy 
between 2\farcs2 (350~au) and 4\farcs7 (750~au).   
Assuming a dust opacity, $\kappa_\nu = 5$~g~cm$^{-2}$, 
and a conservative dust temperature, $T_\mathrm{dust}=30$~K \citep[see, e.g.,][]{beckwith90,andrews11}, 
gives a total unseen dust mass, $M_\mathrm{dust}^\mathrm{out} < 2.5 \times 10^{28}$~g.  
Assuming that the column density is constant over this outer annulus gives an upper limit to 
the dust column density of $\Sigma_\mathrm{dust}^\mathrm{out} \approx 8 \times 10^{-5}$~g~cm$^{-2}$.  
Given that it is likely that the dust emission continues 
to decrease as a function of radius toward the outer disk, 
this is a very conservative upper limit.  
For the canonical gas-to-dust mass ratio of $\sim 100$, this 
implies a gas mass column density of 
$\Sigma_\mathrm{gas}^\mathrm{out} < 8 \times 10^{-3}$~g~cm$^{-2}$ 
(assuming integration in the vertical direction orthogonal to the disk midplane) 
which corresponds to a gas number column density of 
$\lesssim 2 \times 10^{21}$~cm$^{-2}$
for a mean molecular mass of 2.2~amu (i.e., 90\% \ce{H2} and 
10\% He).  
This gas column density is approaching that required for efficient 
self and mutual shielding of CO from photodissociation by photons from the 
central star and/or interstellar medium \citep[see, e.g.,][]{visser09}.  
The simple ``back-of-the-envelope" calculation done 
here implies that the gas-to-dust mass ratio in the outer disk 
is {\em at least} the canonical value of $\sim 100$ to allow 
CO to survive with a sufficient abundance to large radii.  
That the $^{12}$CO emission also originates from layers high up in the 
disk atmosphere (see the next section) provides additional evidence 
that the gas-to-dust mass ratio in the outer regions is likely significantly 
higher than the canonical value, because CO is already able to reach its 
shielding column at large geometrical heights in the disk.  
Given that the temperature structure of the outermost regions of this disk 
is not yet well constrained, it is possible that this emission morphology can also 
be attributed to CO freeze-out in the disk midplane. 
However, we do not see a clear decrement in emission between the front and 
back side of the disk (due to a lack of CO emission from the midplane, 
see Fig.~\ref{figure8}), which has been seen for disks which do possess 
this morphology due to CO freeze-out, e.g., HD~163296 \citep{degregoriomonsalvo13}.
Spatially resolved observations of optically thin tracers will 
confirm which scenario holds.

\begin{figure}
\centering
\includegraphics[width=0.5\textwidth]{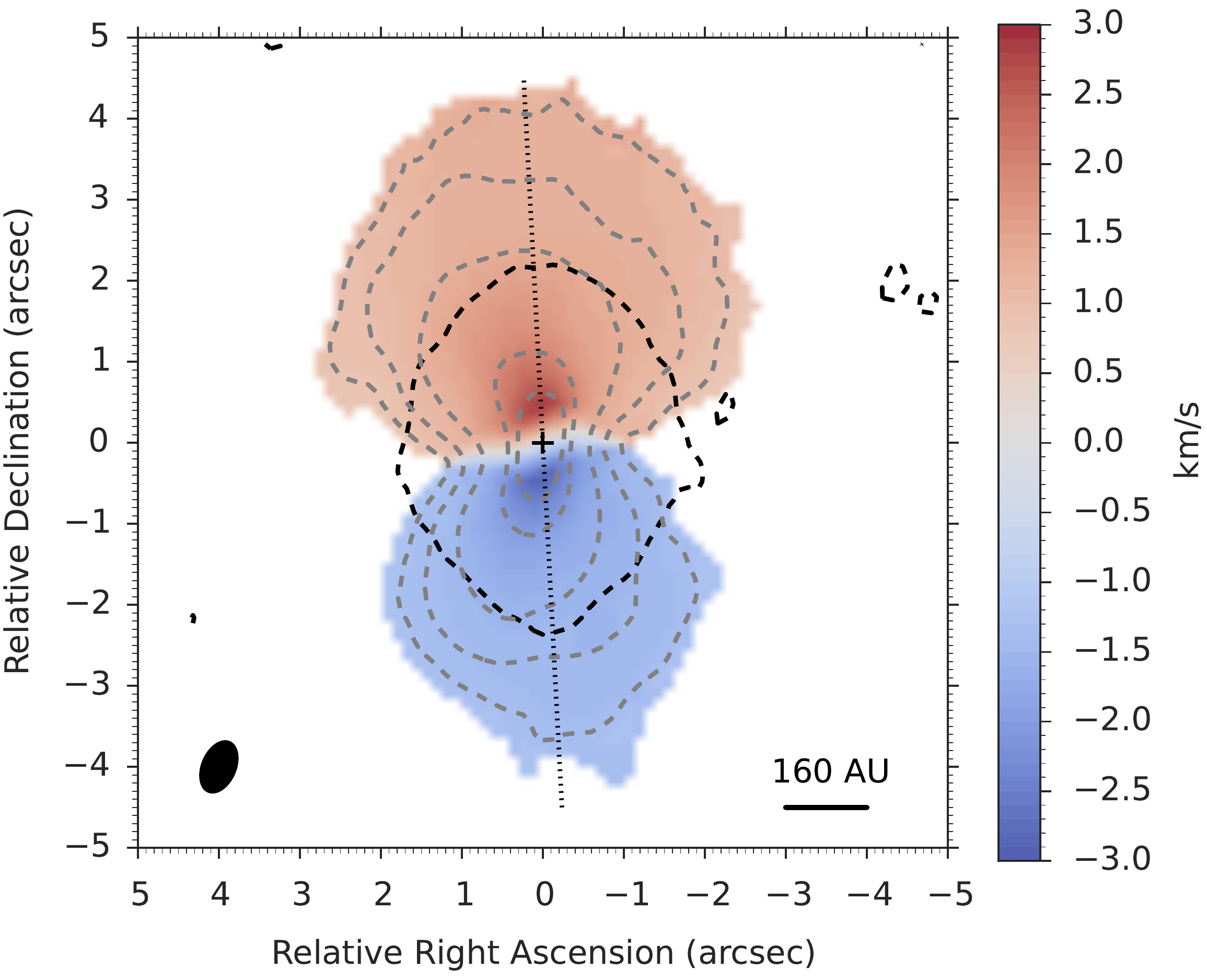}
\caption{CO $J=3$-2 (345.796~GHz) first moment map (color map) and integrated intensity contours 
(dashed gray lines) compared
with the $3\sigma$ contour for the continuum at 346~GHz (dashed black line). 
The 3, 10, 30, 100, and $150\sigma$ contours are shown for the 
integrated intensity. 
The dotted gray line shows the slice across the disk major axis ($3\degree$).}
\label{figure10}
\end{figure}

\begin{figure}[!h]
\centering
\includegraphics[width=0.5\textwidth]{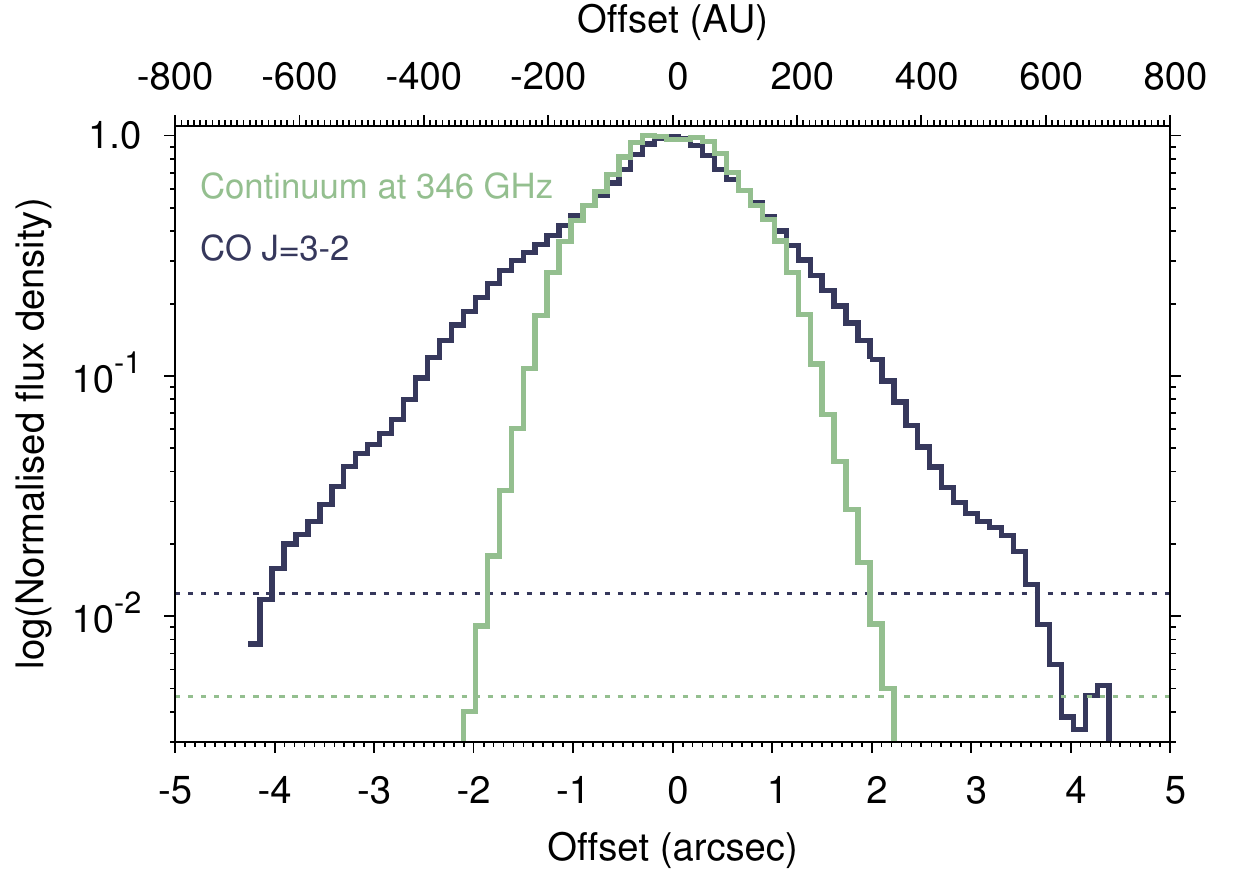}
\caption{Normalized CO $J=3$-2 (345.796~GHz) integrated intensity (dark blue) 
and 346 GHz continuum emission (light green) along the disk major axis.  
The dashed horizontal lines represent the respective $3\sigma$ rms noise levels.}
\label{figure11}
\end{figure}
 
\subsubsection{Constraining the flaring angle of the CO-emitting layer}

The CO $J=3$-2 channel maps (Figure~\ref{figure8}) suggest that the CO emission arises 
from a flared disk.  
For inclined and flared disks, the back side of the disk can become visible in the channel maps: 
this is possible because the line-of-sight projected velocities at the same radius 
from the front and back side will be offset from each other 
\citep[see, e.g.][]{semenov07,degregoriomonsalvo13,rosenfeld13}.  
The back side of the disk is potentially visible in channels from 0.90 to 1.65~km~s$^{-1}$ which 
also show that the CO emission arises from a layer at a particularly large 
geometrical height.  

As a first attempt to constrain the opening angle (relative to the midplane) 
of the CO-emitting layer, we plot the channel maps mirroring the positive and 
negative channels at the same velocity (from $\pm0.75$ to $\pm1.80$~km~s$^{-1}$, Figure~\ref{figure12}).  
The line emission arises from a layer clearly angled toward the east (in the 
direction of increasing R.A.) relative to that expected 
for an inclined geometrically flat disk.  
Also overlaid on the mirrored channel maps are the projected angles of a layer inclined 
at 41$\degree$ relative to the disk P.A. (3$\degree$) for an opening angle 
of 30$\degree$, 35$\degree$, and 40$\degree$ 
(white dotted lines, with the angle increasing from right to left).  
Further refinement of the opening angle without detailed modeling 
(such as that conducted by \citet{rosenfeld13} and beyond the scope of this work) 
is difficult because of the lack of emission 
in and around the source velocity. 
This simple geometrical analysis shows that the opening angle of CO-emitting layer is 
$\approx35\pm5\degree$ confirming the flared nature of this protoplanetary disk.  

In the previous section, we derived the radius of the molecular disk by assuming that the 
emission arises from a geometrically thin disk and modeling the CO $J=3$-2 integrated intensity 
in the $uv$ plane.  
That the CO arises from a layer which is significantly inclined with respect to the 
midplane should not affect the determination of the disk radius by this method, since 
the projected radius of the flared layer will coincide with the radial extent of 
the disk along the midplane. 
 
\begin{figure*}[!h]
\includegraphics[width=\textwidth]{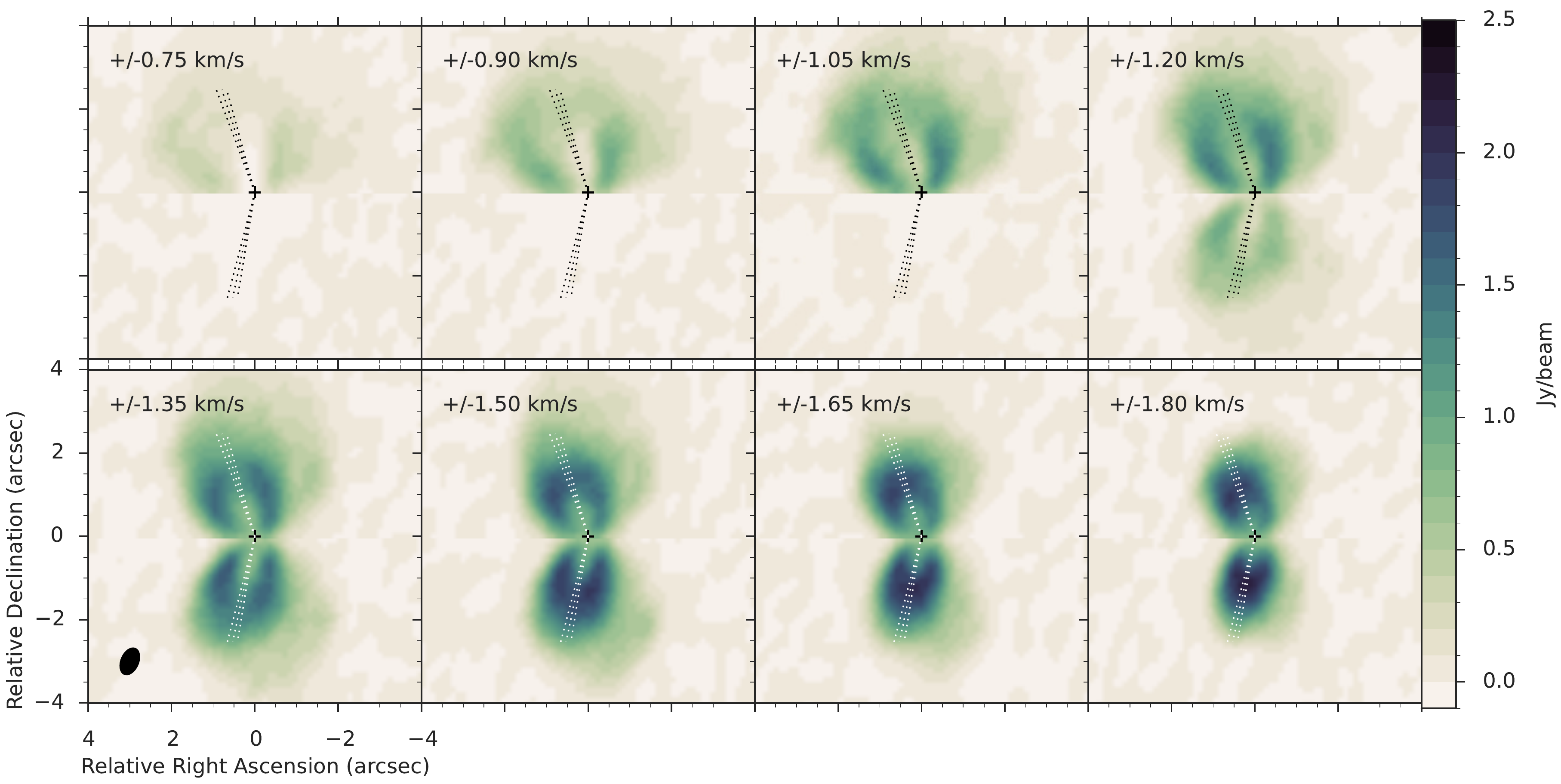}
\caption{Channel maps showing the CO J=3-2 (345.796~GHz) line emission 
with the positive and negative channels mirrored across zero relative declination.  
The black cross indicates the source position and the synthesized beam 
is shown in the bottom left-hand corner of the bottom left-hand panel.  
The dotted white and black lines show the projected angle of a surface inclined 
at $41\degree$ relative to the disk P.A. $3\degree$ with an opening angle (relative to 
the disk midplane) of 30$\degree$, 35$\degree$, and 40$\degree$ (increasing from right to left).}
\label{figure12}
\end{figure*}

\section{DISCUSSION}
\label{discussion}

\subsection{On the transitional nature of HD~97048}

The data presented here support the proposed transitional nature of the 
disk encompassing HD~97048, with the data and extracted intensity profile 
showing a decrement in continuum emission within $\approx50$~au. 
The data also show that the molecular gas and small ($\mu$m-sized) 
dust grains have a larger radial extent than the large (mm-sized) dust grains 
(750~au versus 350~au).  
Better data are needed to quantify the dust depletion factor within the 
identified cavity.   
This discrepancy in radial extent between the (sub-)mm-sized dust and molecular gas may be 
indicative of dust evolution during the disk 
lifetime: as dust grains grow, they become decoupled from the gas, feel a 
headwind from the slightly sub-Keplerian gas, and move inwards to conserve 
angular momentum, a process termed {\em radial drift} 
\citep[see, e.g.,][]{whipple72,weidenschilling77}.  
This has been seen in several other disks imaged at 
(sub-)mm wavelengths at relatively high spatial resolution ($\lesssim$0\farcs5), 
including TW~Hya \citep{andrews12,andrews16,hogerheijde16}, LkCa~15 \citep{isella12}, 
HD~163296 \citep{isella09,degregoriomonsalvo13}, and 
HD~100546 \citep{pineda14,walsh14}.  
Sharply truncated dust disks are predicted by dust evolution models 
which simulate radial drift coupled with viscous gas drag \citep{birnstiel14}. 
\citet{birnstiel14} predict that the radius of (sub-)mm dust disks
can decrease to values as low as 1/4--1/3 that of the gaseous disk by 
$10^{6}$ years.  
Such ``sharp'' outer edges are found to form quickly during 
early stages in the disk lifetime ($\sim10^{5}$~year) and 
can persist to late stages ($\sim10^{6}$~year) if the drift becomes 
``self-limited,'' i.e., as dust is lost to the star, the dust-to-gas mass ratio 
decreases, and the maximum achievable particle size reduces \citep{birnstiel14}.  
HD~97048 is considered a rather young source, with 
an estimated age of $\approx2-3$~Myr 
\citep{vandenancker98,lagage06,doering07,martin-zaidi09}.  
The ALMA data presented here suggest a ratio for the radius of 
(sub-)mm-sized dust grains to that of the molecular gas of $\lesssim0.5$, if we take the 
radial extent of the continuum and CO line emission at face value (see the discussion in Sect.~\ref{diskradius}).  
This is 
slightly larger than that predicted by the dust evolution models at $\sim10^{6}$~year 
\citep{birnstiel14}.  

The presence of the inner cavity in the large dust grains in HD~97048 
is indicative that some physical mechanism is in operation which is impeding 
radial drift and thus maintaining the population of large dust grains out to 
several hundred~au.
\citet{quanz12} do not discuss any evidence of gaps and substructure in the small grain 
populations in existing PDI images of HD~97048. If confirmed, this likely rules out 
photoevaporation as the origin of the cavity in the larger dust grains;  
however, \citet{vanderplas09} do infer a cavity in the CO gas within the innermost 11~au using 
high-spectral-resolution vibrational line emission.  
That [OI] emission has been detected in the inner region \citep[$0.8-20$~au,][]{acke06}, 
further supports the idea that CO gas is photodissociated by the strong far-UV flux from 
the star on small scales.  
Alternative theories include the presence of a planet or planetary system inside the 
cavity which creates a pressure gradient in the gas and 
traps the dust in a ring-like structure external to the location of the planet(s) 
\citep[see, e.g.,][]{pinilla12b}.  
The extracted intensity profile for HD~97048 also shows the presence 
of rings (at $\approx50$, 150, and 300~au) with 
associated gaps (at $\approx100$ and 250~au). 
If such substructure were caused by forming planets or companions {\em within} the 
protoplanetary disk, then this would also help to maintain a population of 
large dust grains out to several hundred au and increase the ratio of the radii 
of the (sub-)mm-sized dust and molecular gas disks relative to that predicted 
by models which do not include the influence of planets.  
A recent reanalysis of ALMA Cycle 0 continuum emission from the disk around TW~Hya 
\citep{hogerheijde16} predicted the presence of 
one or more unseen embedded planets to explain the radial extent of 
(sub-)mm-sized dust grains; recent high (0\farcs3) to very high (0\farcs02) 
angular-resolution images of continuum emission from TW Hya confirm the presence of 
multiple rings and gaps which could be carved by the postulated unseen embedded 
planets \citep{andrews16,nomura16,tsukagoshi16}.

\subsection{HD~97048 versus HD~100546: both planet-hosting disks?}

HD~97048 and HD~100546 are both disk-hosting stars with a very similar spectral type.  
Multiple dust rings have been observed in the disk around 
HD~100546 with ALMA \citep{walsh14,pinilla15}, attributed to the presence of two massive companions:  
one orbiting within the cavity \citep[$<10$~au, e.g.,][]{acke06,brittain14,currie15} and one, 
likely very young object, directly imaged at $\approx50$~au \citep{currie14,currie15,quanz15}.  
The extracted intensity profile for HD~97048 also shows 
evidence of substructure with peaks in emission at 
$\approx50$, 150, and 300~au and gaps in emission at $\approx100$ and 250~au.  
The cavity size for HD~97048 as determined from these data is $\approx25$~au 
which corresponds to the innermost radius of the FWHM  
of the innermost peak in emission (see Figure~\ref{figure6}).  
We note here that the high signal-to-noise of the data ($\sim 1000$) coupled with 
data analysis conducted in the visibility domain allows the extraction 
of substructure from the interferometric data on size scales which are smoothed 
over significantly in the resulting images.

In contrast with HD~100546, no observational evidence of a massive inner companion 
around HD~97048 yet exists, either in the disk continuum observations, or 
in the disk gas observations. 
However, elemental abundance measurements 
in the photosphere of HD~97048 hint that the gas-to-dust ratio of 
accreting material is large ($750\pm250$) compared with the canonical interstellar medium  
value of $\sim100$ \citep{acke04,kama15}.  
\citet{kama15} hypothesize that this is a potential indirect determination 
of the presence of unseen inner planets/companions in Group I disks: the presence of planets 
in an inner gap or cavity impedes the flow of dust through the gaps whereas gas can flow 
freely, thereby locally increasing the gas-to-dust ratio in the accreting zone.   
The presence of an inner companion with an appreciable mass within the dust cavity 
will also excite spiral density waves which may be observable with 
future high-spatial observations at optical wavelengths.

Assuming that a single planet is responsible for clearing the inner cavity and trapping the 
(sub-)mm-sized dust beyond a radius of 25~au, 
one can estimate the mass and location of the planet 
\citep[see, e.g, ][]{crida06,dodsonrobinson11,pinilla12b}.  
\citet{pinilla12b} find that the dust accumulates at $\approx7 r_{H}$ for planets 
with masses of $1-3 M_\mathrm{Jup}$ and at $\approx10 r_\mathrm{H}$ for planets 
with mass $> 5M_\mathrm{Jup}$, where $r_\mathrm{H}$ is the Hill radius 
($r_\mathrm{H} = r_p(M_p/3M_\star)^{1/3}$, where $r_p$ and $M_p$ are the radial location and mass of the 
planet and $M_\star$ is the stellar mass).  
In Figure~\ref{figure13}, the light-green shaded region highlights the range of possible 
planet orbital locations and masses for a sub-mm dust ring located 
at radii between 25 and 50~au  
(as suggested in the ALMA data).  
Also plotted are the radial ranges of the optically thick inner disk 
\citep[shaded gray region, $0.3-2.5$~au,][]{maaskant13}, 
the cavity size inferred from the ALMA data (light-green striped region, $25-50$~au, this work), 
and that inferred from the mid-IR imaging data 
\citep[dark-blue striped region, $30-38$~au,][]{maaskant13}. 
The dark-blue shaded region highlights the range of values when a cavity 
size of $34\pm4$~au is assumed (as inferred from the mid-IR data).  
The range of planet locations and masses is set by the size of the 
gap carved in the gas \citep[$\approx 5 r_H$,][]{dodsonrobinson11} and the maximum dust gap size seen 
in the dust evolution models \citep[$\approx 10 r_H$,][]{pinilla12b}.  

This simple parameterization shows that a $1~M_\mathrm{Jup}$ planet
is able to create a dust trap at the required radius if it is located 
at  $29\pm12$~au for a dust trap located between 25 and 50~au and 
$25\pm5$~au for a dust trap at $34\pm4$~au.  
However, such a low-mass planet is not able to open up a gap as wide as 
the full cavity inferred from both the ALMA data and the mid-IR imaging.  
It is more likely that a single planet has a mass 
$>10~M_\mathrm{Jup}$, with planet/companion masses on the order of $100~M_\mathrm{Jup}$ 
also possible.  
To date, no massive companion has been inferred from existing data on HD~97048,  
and no constraints on the mass of an as-yet unseen planet have been been determined
\citep[see, e.g.,][]{acke06,vanderplas09,brittain14}. 
That the cavity inferred in HD~97048 is larger than that for HD~100546, 
despite the lack of constraints on the presence or otherwise of a massive 
inner companion, is also intriguing. 
Very recently, \citet{dong15} presented results from 2D hydrodynamical models in which 
they show that multiple (four) low-mass ($1-2M_\mathrm{Jup}$) planets can open gaps on the 
order of a few tens of au in the mm-sized dust grains.

HD~97048 is yet another disk for which the dust emission at (sub-)mm wavelengths 
shows evidence of axisymmetric ring-like structures, here on spatial
scales of around tens of au \citep{walsh14,alma15,andrews16,nomura16,zhang16}.  
We predict that this substructure will be clearly evident in images of HD~97048 at higher 
spatial resolution ($\approx10-20$~au, see Figure~\ref{figure14}).
There remains much debate in the literature on the origin of such axisymmetric 
substructure in protoplanetary disks including gaps and dust traps carved by forming 
planets \citep[see, e.g.,][]{dipierro15,pinilla15,rosotti16}, a change in dust opacity properties at the 
positions of snow lines \citep[e.g.,][]{banzatti15,zhang15,guidi16,okuzumi16}, 
and toroidal dust traps created by hydrodynamic or magnetohydrodynamic effects 
\citep[see, e.g.,][]{pinilla12a,lorenaguilar15,ruge16}.  
To distinguish between each of the scenarios requires observations of 
dust emission at multiple and well-separated frequencies to determine the 
radial dust size and density distribution (and dust opacity index) along with 
emission from optically thin gas tracers to determine the gas surface density.  
Planets will create deep gaps in the gas surface density as well as influencing the 
dust \citep[note that this is dependent on the planet mass, see e.g.,][]{rosotti16}, 
toroidal instabilities will create much shallower features in the gas surface 
density, and opacity changes at snow lines will affect only the dust emission and will 
have no effect on the gas.  
We note that the ringed substructure seen here has very recently been confirmed in 
scattered light images of HD~97048 taken with VLT/SPHERE \citep{ginski16}.  
An initial (and shallow) comparison of the data sets shows remarkable coincidence 
between the positions of the (sub-)mm peaks and gaps and those seen in scattered light.  
That such structure is seen in both small ($\approx~\mu$m-sized) dust grains 
in the disk atmosphere and large ($\approx$~mm-sized) dust grains in the disk midplane 
points toward a (proto)planetary system origin; however, further data, particularly 
to better constrain the gas structure, are needed for confirmation. 
Since this paper has been accepted for publication, ALMA Cycle 2 data of HD~97048, 
for which longer baseline data were available and imaged with a $uv$ clip ($> 160 k\lambda$), 
resulted in a beam of $0\farcs48 \times 0\farcs26$ (18\degree) 
and resolved the inner dust cavity ($< 40-46$ au) 
and the bright dust ring at $\approx150$~au \citep{vanderplas16}. 
  
\begin{figure}[!h]
\centering
\includegraphics[width=0.5\textwidth]{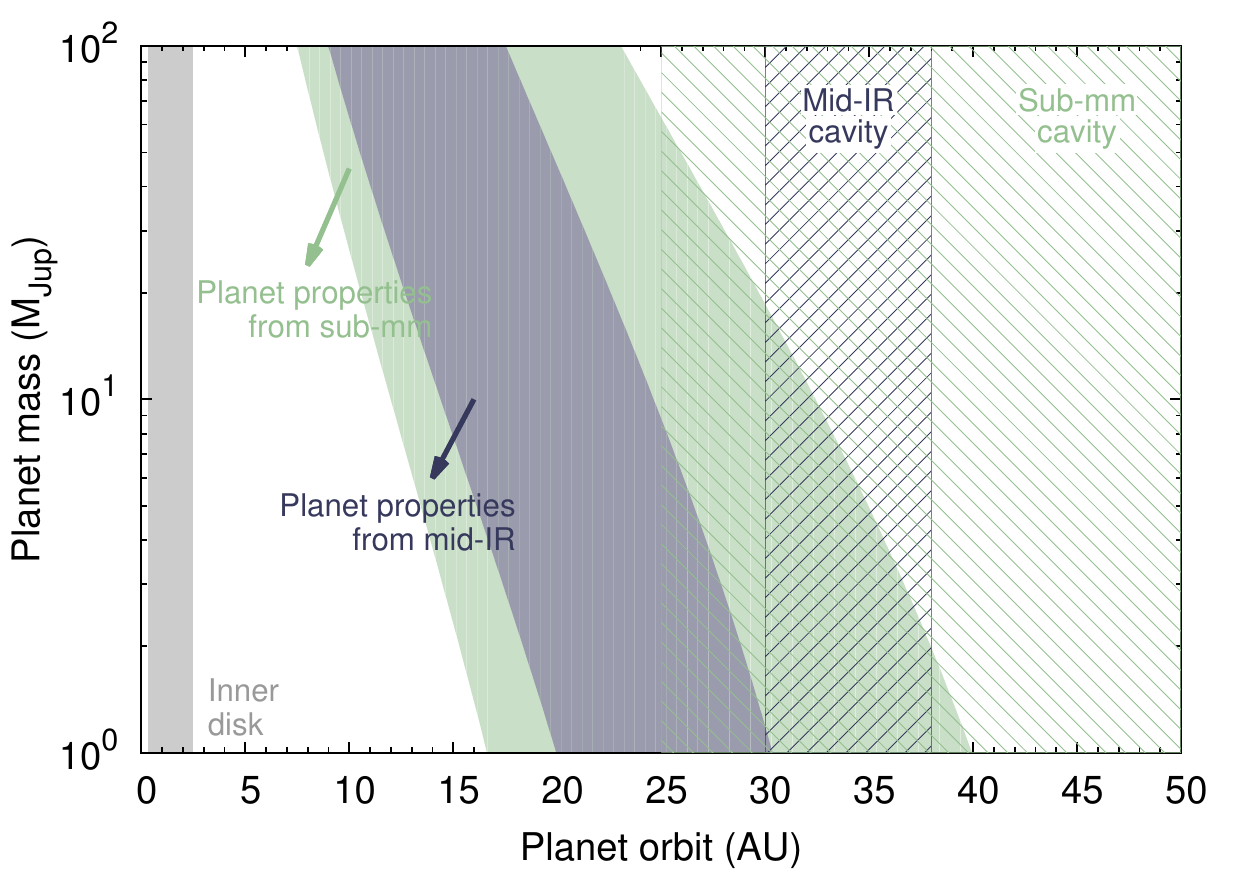}
\caption{Plot showing the range of possible planet locations and masses 
for a single planet responsible for creating a dust trap between 25 and 50~au
(light-green shaded area), and at $34\pm4$~au (dark-blue-shaded area).  
Also shown are the size of the optically thick inner dust disk 
\citep[gray box,][]{maaskant13}, the cavity radius inferred from the ALMA data 
(light-green striped region, this work), and that inferred from mid-IR imaging 
\citep[dark-blue striped region,][]{maaskant13}.}
\label{figure13}
\end{figure}

\begin{figure}[!h]
\includegraphics[width=0.5\textwidth]{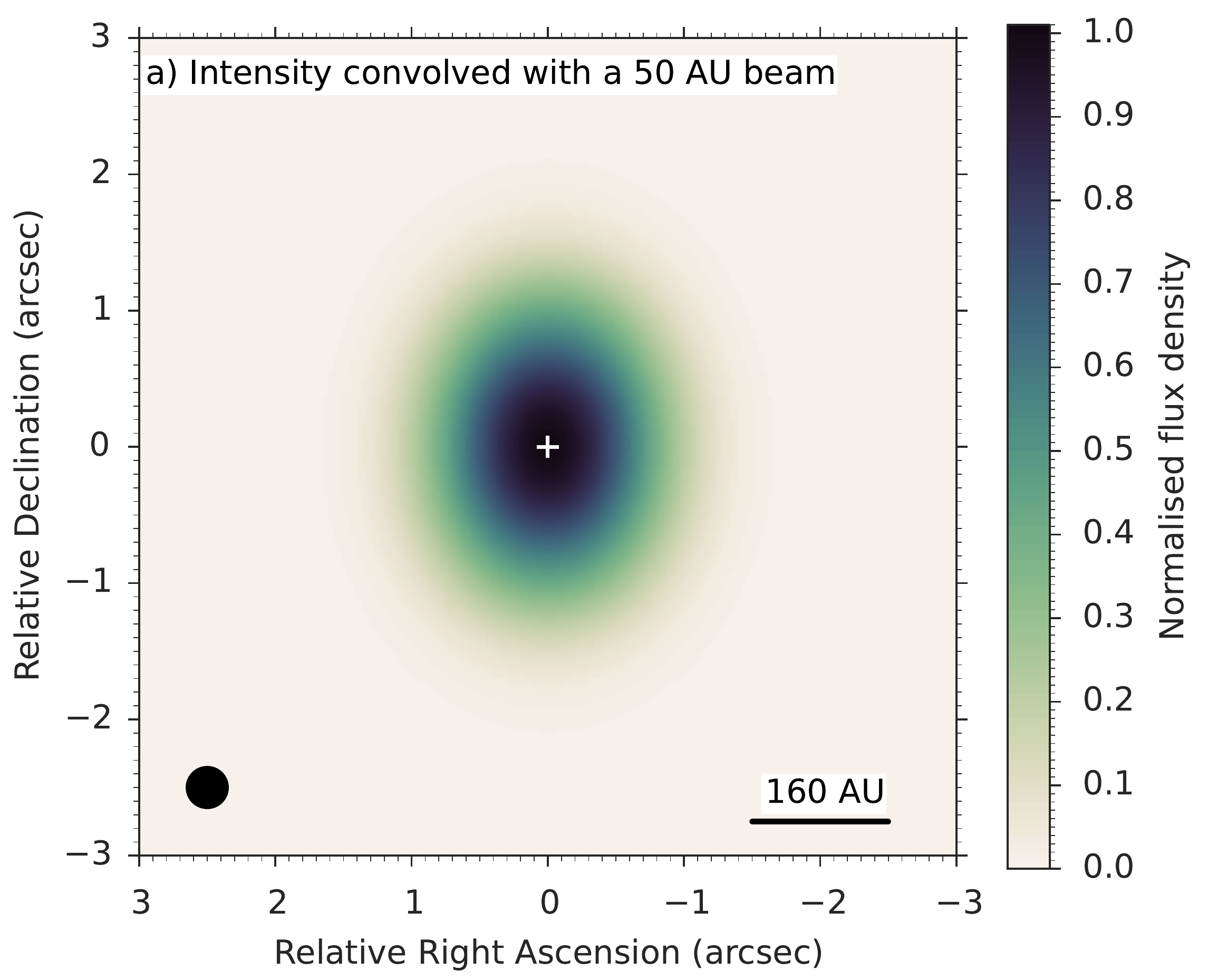}
\includegraphics[width=0.5\textwidth]{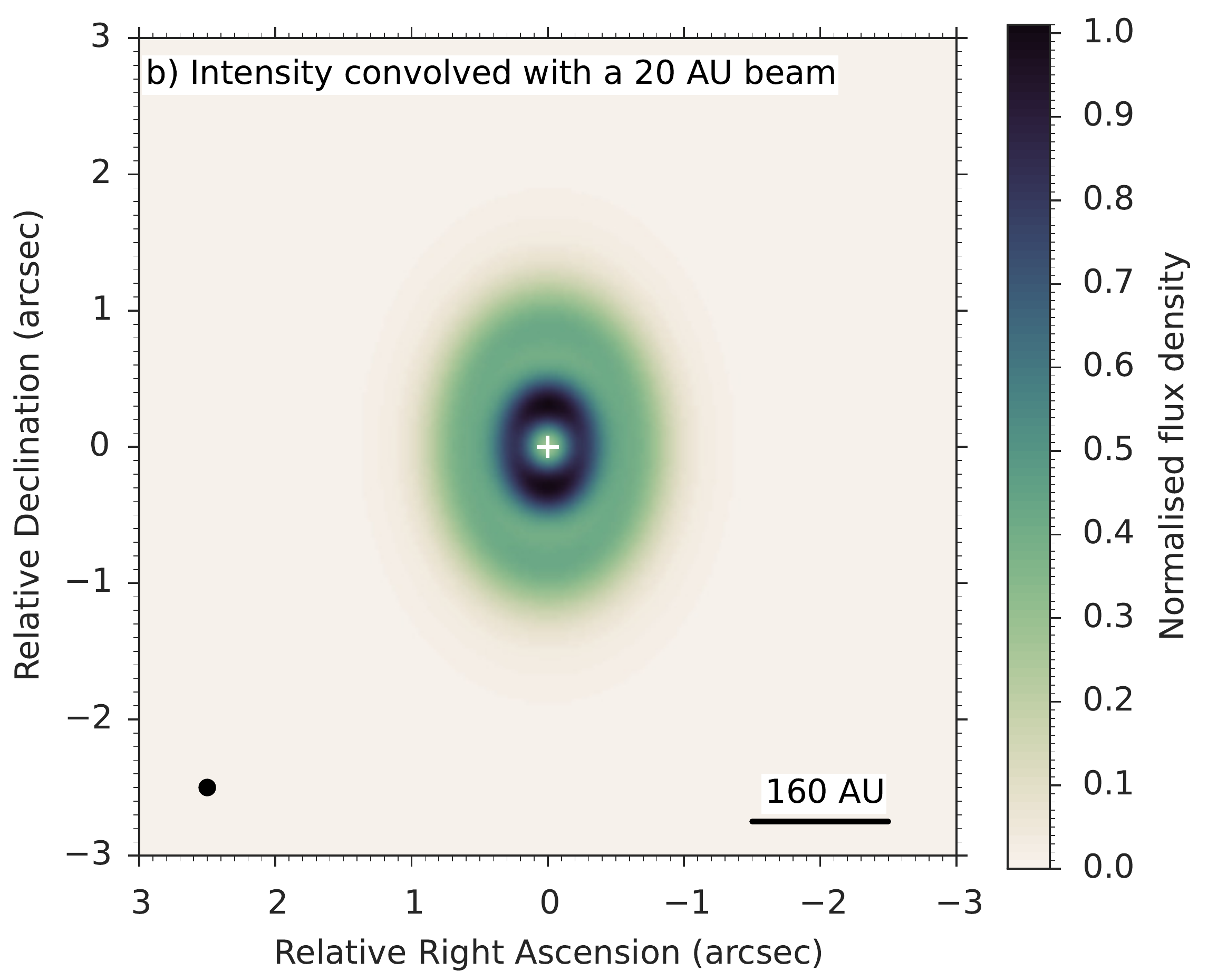}
\caption{Normalized flux density using the best-fit intensity profile convolved 
with a 50~au beam (left-hand panel) and a 20~au beam (right-hand panel).}
\label{figure14}
\end{figure}

\subsection{HD~97048: an extremely flared disk?}
The CO $J=3$-2 data show that the molecular disk around HD~97048 extends to at 
least the same radius as the small dust grains imaged in scattered light 
\citep{doering07}, confirming that the scattered light does arise 
from a large circumstellar disk as opposed to a remnant circumstellar 
envelope.  
The CO data suffer from missing emission at and around the source velocity which could 
be due to foreground absorption and/or spatial filtering of background cloud emission; 
however, the data quality is sufficient to constrain the source 
velocity ($V_\mathrm{LSRK}$) to $4.65\pm0.075$~km~s$^{-1}$. 
As discussed in Sect.~\ref{hd97048}, there exist limited data  
on emission from the outer regions of the molecular disk: 
most data are spatially and spectrally unresolved and probe the 
inner warm/hot surface layers \citep{meeus12,meeus13,fedele13,vanderwiel14}.   
\citet{hales14} report the detection of CO $J=3$-2 towards HD~97048 with APEX 
but identify the emission as arising from background material,  
because of an apparent offset in source velocity.  
Correcting the source velocity to 4.65~km~s$^{-1}$ in figure~18 in \citet{hales14} 
shows that the emission does indeed arise from a disk in Keplerian 
rotation, albeit with a red wing which is significantly stronger than the blue wing 
and also with significant missing emission at the source velocity 
(as also seen in this work).  
  
The data also show that the CO $J=3$-2 emission arises from a layer
with an opening angle (relative to the midplane) of $\approx30\degree-40\degree$.  
This translates to a ratio, $z/r\approx0.6-0.8$ (where $z$ is the geometrical height and 
$r$ is the disk radius), identifying HD~97048 as one of the most flared 
HAeBe disks known to date in $^{12}$CO line emission, which is 
consistent with its SED classification as a (flared) Group I protoplanetary disk 
\citep{meeus01,acke10}.  
In comparison, \citet{rosenfeld13} derive an opening angle 
of $\approx15\degree$ for the $^{12}$CO-emitting surface in the 
disk around HD~163296 \citep[similar to that also derived by][]{degregoriomonsalvo13}. 
This corresponds to $z/r\approx0.3$ and is also consistent 
with its SED classification as a (flat) Group II protoplanetary disk \citep{meeus01}.  
Interestingly, \citet{bruderer12} reproduce the single-dish observations 
toward the Group~I disk, HD~100546, with a model which suggests 
that the CO $J=3$-2 emission arises from a layer deeper than that for HD~163296, 
at $z/r\approx0.2$.  
\citet{lagage06} fit an intensity profile to the $8.6~\mu$m 
PAH emission from HD~97048 using a power law, $z=(z_0/r_0)r^{\beta}$, where 
$z_0$ is the geometrical height of the emitting layer at 
$r_0$ and $\beta$ is the power-law index \citep[see also][]{chiang01}.  
They derive the following parameters: $z_0=51.3$~au and $\beta=1.26$, for 
a fixed $z_0=135$~au.  
This translates to $z/r=1.7$ at 300~au (roughly the spatial extent 
of the $8.6~\mu$m PAH emission), confirming that the CO J=3-2 emission 
arises from layer deeper in the disk than the far-UV-excited PAHs, 
as would be expected.  
Note that the PAH emission and CO $J=3$-2 line emission both arise from 
layers in the disk at geometrical heights several times that of the expected 
gas pressure scale height.

\subsection{Future outlook}
The data presented here are the first to spatially resolve the 
large disk around HD~97048 at (sub-)mm wavelengths.  
The data confirm the presence of a cavity in the large dust grains; 
however, higher spatial resolution data, ideally at multiple frequencies, 
are required to determine the depth of the gap in both dust and molecular gas  
and thus constrain the physical mechanism which is halting radial drift in 
the inner disk.  
High spatial resolution data will also shed light on the possible presence 
of a multiple planetary system composed of $\approx$ Jupiter-sized planets 
and will also confirm the multiple rings seen in the continuum emission 
in the current data.  
Finally, spatially resolved observations of multiple transitions of optically thin 
CO isotopologues will help to further constrain the location of the 
flared CO-emitting layer, and allow derivation of the temperature structure of 
the molecular gas. 
Quantification of the gas surface density will also allow further distinction 
between the various theories for the creation of the ring-like structures seen 
in the dust emission.
 
\acknowledgements
This paper makes use of the following ALMA data: ADS/JAO.ALMA\#2011.0.00863.S. 
ALMA is a partnership of ESO (representing its member states), 
NSF (USA) and NINS (Japan), together with NRC (Canada) and NSC and 
ASIAA (Taiwan), in cooperation with the Republic of Chile. 
The Joint ALMA Observatory is operated by ESO, AUI/NRAO and NAOJ.  
C.~W. is supported by the Netherlands Organisation for Scientific Research (program number 639.041.335).  
A.~J. is supported by the DISCSIM project, grant agreement 341137 funded by the European Research Council under
ERC-2013-ADG.
G.~M. is supported by Spanish grants RYC-2011-07920 and (partly) AYA2014-55840-P.
Astrophysics at QUB is supported by a grant from the STFC.  
The authors thank Drs.~N. van~der~Marel, M.~Kama, P.~Pinilla, M.~Hogerheijde, 
E. Chapillon, and G.~Sandell for useful discussions, and an anonymous referee
for providing constructive comments which improved the discussion in the paper.

\end{document}